\documentclass[12pt,doublespacing]{article}
\usepackage{fullpage}
\usepackage{setspace}
\usepackage{authblk}
\usepackage{cite}
\usepackage[title]{appendix}
\usepackage{amssymb}
\usepackage{graphicx}
\usepackage{graphics}
\usepackage[cmex10]{amsmath}
\usepackage{latexsym,amsfonts}
\usepackage{amsthm}
\usepackage{subfigure}
\usepackage{hyperref} 
\usepackage{enumitem}

\begin{document}

\title{A Coupled Hybridizable Discontinuous Galerkin and Boundary Integral Method for Analyzing Electromagnetic Scattering}
\author[1]{Ran Zhao}
\author[2]{Ming Dong}
\author[2]{Liang Chen} 
\author[3]{Jun Hu} 
\author[2]{Hakan Bagci} 

\affil[1]{Ran Zhao is with the Division of Computer, Electrical, and Mathematical Science and Engineering,
King Abdullah University of Science and Technology (KAUST), Thuwal 23955, Saudi Arabia, and also with
the School of Electronic Science and Engineering, University of Electronic Science and Technology of China
(UESTC), Chengdu 611731, China (e-mail: ran.zhao@kaust.edu.sa).\vspace{0.25cm}}
\affil[2]{Ming Dong, Liang Chen, and Hakan Bagci are with the Division of Computer, Electrical, and Mathematical
Science and Engineering, King Abdullah University of Science and Technology (KAUST), Thuwal 23955,
Saudi Arabia (e-mail: {ming.dong, liang.chen, hakan.bagci}@kaust.edu.sa).\vspace{0.25cm}}
\affil[3]{Jun Hu is with School of Electronic Science and Engineering, University of Electronic Science and Tech-
nology of China (UESTC), Chengdu 611731, China (e-mail: hujun@uestc.edu.cn).}
\footnotetext[1]{This work was supported in part by the National Natural Science Foundation of China under Grant 62271002 and Grant 62031010, and in part by KAUST OSR under Award 2019-CRG8-4056.}
\date{}
\maketitle

\newpage

\begin{abstract}
A coupled hybridizable discontinuous Galerkin (HDG) and boundary integral (BI) method is proposed to efficiently analyze electromagnetic scattering from inhomogeneous/composite objects. The coupling between the HDG and the BI equations is realized using the numerical flux operating on the equivalent current and the global unknown of the HDG. This approach yields sparse coupling matrices upon discretization. Inclusion of the BI equation ensures that the only error in enforcing the radiation conditions is the discretization. However, the discretization of this equation yields a dense matrix, which prohibits the use of a direct matrix solver on the overall coupled system as often done with traditional HDG schemes. To overcome this bottleneck, a ``hybrid'' method is developed. This method uses an iterative scheme to solve the overall coupled system but within the matrix-vector multiplication subroutine of the iterations, the inverse of the HDG matrix is efficiently accounted for using a sparse direct matrix solver. The same subroutine also uses the multilevel fast multipole algorithm to accelerate the multiplication of the guess vector with the dense BI matrix. The numerical results demonstrate the accuracy, the efficiency, and the applicability of the proposed HDG-BI solver.\par
\end{abstract}
\newpage

\section{Introduction}
In the past two decades, the discontinuous Galerkin (DG) method~\cite{hesthaven2007nodal,gedney2012discontinuous,bao2019pml,lu2004discontinuous,chen2012discontinuous,Gedney2009,ren2016eb,Li2015resistive,alvarez2015efficient,zhan2020stabilized,yan2017discontinuous,wang2019parallel,chen2020multiphysics,chen2020memory} has attracted significant attention in the computational electromagnetics research community because, compared the traditional finite element method (FEM)~\cite{Jin2015FEM,lee1995whitney,lee1997time,jiao2002general,he2012explicit}, it offers a higher-level of flexibility in discretization which allows for non-conformal meshes and an easier implementation of h-and/or p-adaptivity. In addition, in time domain, when combined with an explicit time integration scheme, the DG method produces a very compact, fast, and easy-to-parallelize solver since the DG's block diagonal mass matrix is inverted once and very efficiently before the time marching is started. However, this increased efficiency does not carry over to the frequency domain. Due to doubling of the unknowns at the element boundaries and the fact that a sparse matrix system must still be solved, frequency-domain DG schemes usually require more computational resources than the traditional frequency-domain FEM.

Recently, this drawback has been alleviated with the introduction of the hybridizable discontinuous Galerkin (HDG) method~\cite{cockburn2009unified}. HDG introduces single-valued hybrid variables on the skeleton of the mesh (namely, a mesh that consists of only the faces of the elements)~\cite{cockburn2009unified} and converts the local/elemental DG matrix systems into a coupled global matrix system, where these hybrid variables are the unknowns to be solved for. The computational requirements of HDG are lower than DG since the total  degrees of freedom is now reduced~\cite{cockburn2009unified}.

Indeed, HDG is competitive to FEM in terms of computational requirements when both methods use a high-order discretization, and at the same time, it maintains the advantages of the traditional DG over FEM~\cite{kirby2012cg,yakovlev2016}. In addition, thanks to the local post-processing used after the global matrix system solution, HDG achieves an accuracy convergence of order $p + 2$ (superconvergence), where $p$ is the order of polynomial basis functions used to expand the local/elemental field variables~\cite{cockburn2016static}. Because of these benefits, HDG has been used to solve various equations of physics, such as convection-diffusion equations~\cite{nguyen2009an}, Poisson equation~\cite{chen2020hybridizable}, and elastic/acoustic wave equations~\cite{nguyen2011high}. For electromagnetics, HDG was first used to solve the two-dimensional (2D) Maxwell's equations~\cite{nguyen2011hybridizable}. Since then, it has been extended to solve the three-dimensional (3D) Maxwell's equations and used in conjunction with a Schwarz-type domain decomposition method to analyze electromagnetic scattering from large objects~\cite{li2013hybridizable,li2014hybridizable}. In addition, a hybridizable discontinuous Galerkin time-domain method (HDGTD) has been proposed to solve the time-dependent Maxwell's equations. This method combines an implicit and explicit time integration scheme and HDG for time marching and spatial discretization, respectively.~\cite{christophe2018implicit,li2019new,nehmetallah2020explicit}. HDG has also been used in simulations of multiphysics problems: In~\cite{li2017hybridizable,vidal2021nested}, the coupled system of the Maxwell's equations and the hydrodynamic equation has been solved using HDG to simulate the non-local optical response of nanostructures.

Most of the HDG methods, which have been developed to simulate wave interactions, use approximate absorbing boundary conditions (ABCs) to truncate the computation domain~\cite{engquist1977,baylis1980,mur1981,peterson1988,peterson1992}. Although these boundary conditions yield sparse matrices upon discretization, their accuracy is limited and therefore they restrict the high-order convergence of the solution unless a very large computation domain is used. One can also use the method of perfectly matched layer (PML) to truncate the HDG computation domain~\cite{berenger1994,berenger1996,chew1994,chew1997}. Indeed HDG with PML has recently been used in the simulation of waveguide transmission problems~\cite{li2021pml}. However, to increase the ``absorption'' of PML (i.e., to increase its accuracy), one has to increase thickness of the layer or the value of the conductivity. The first option increases the size of the computation domain while the second option has to be done carefully since large values of conductivity often result in numerical reflection from the PML-computation domain interface and decrease the accuracy of the solution~\cite{chen2020memory,Sirenko2012}.
\begin{figure}[t!]
\centerline{\includegraphics[width=0.6\columnwidth,draft=false]{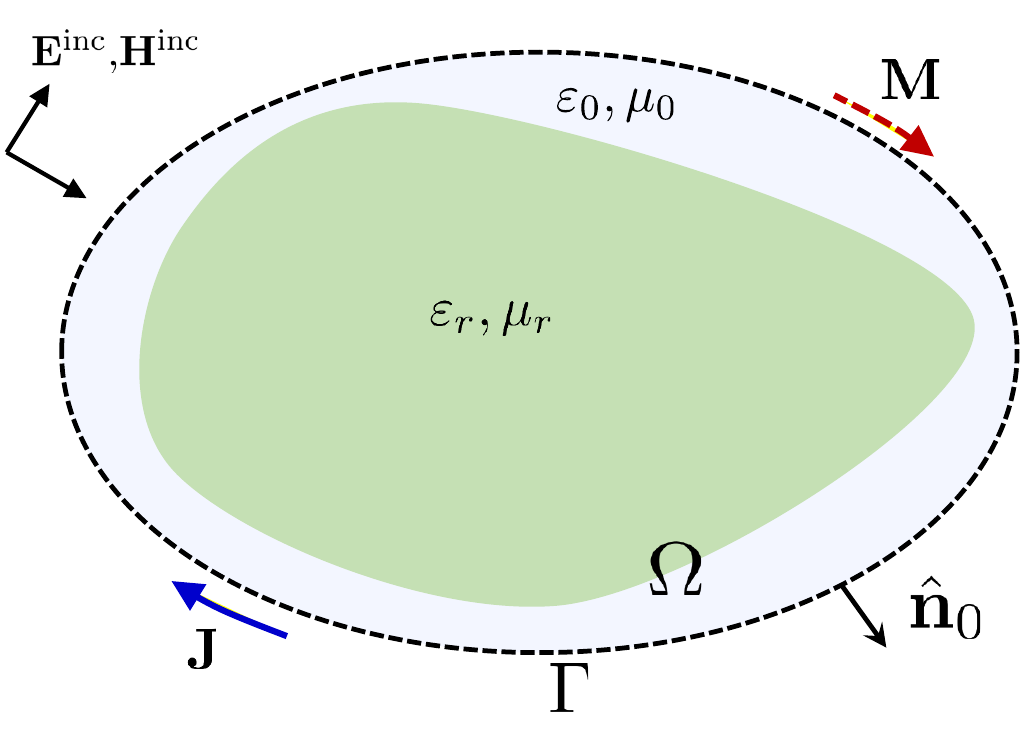}}
\caption{Description of the electromagnetic scattering problem.}
\label{fig:oricomp}
\end{figure}

On the other hand, boundary integral (BI)-based approaches to truncating computation domains do not suffer from these bottlenecks~\cite{jin1991, lu1996,sheng1998formulation,vouvakis2004,botha2004,jiao2002, yilmaz2007, li2014hybrid, dong2022}. In this work, HDG is used together with a BI formulation to efficiently and accurately simulate electromagnetic scattering from electrically large inhomogeneous/composite objects. Since the BI formulation enforces the radiation condition without any approximations, the accuracy of computation domain truncation is only restricted by the discretization error. Furthermore, the surface, where the BI equation is enforced, can be located very close, even conformal, to the surface of the scatterer without any loss of accuracy.

However, the discretization of the BI equation yields a dense matrix, which prohibits the use of a direct matrix solver on the overall coupled system as often done with traditional HDG schemes~\cite{li2013hybridizable,li2014hybridizable}. To overcome this bottleneck, in this work, a ``hybrid'' method is developed. This method uses an iterative scheme to solve the overall coupled system but within the matrix-vector multiplication subroutine of the iterations, the inverse of the HDG matrix is efficiently accounted for using a sparse direct matrix solver. The same subroutine also uses the multilevel fast multipole algorithm (MLFMA)~\cite{rokhlin1990rapid,lu1994multilevel,song1997multilevel,sheng1998solution,fostier2008,ergul2008,michiels2015,yucel2018,abduljabbar2019} to accelerate the multiplication of the guess vector with the dense BI matrix. Another contribution of this work is that it describes in detail the first use of vector basis functions~\cite{nedelec1980mixed} within the HDG framework.

The rest of this paper is organized as follows. Section II first describes the electromagnetic scattering problem and introduces the mesh used to discretize the computation domain. This is followed by the formulation of the coupled HDG and BI equations and the description of the matrix system that is obtained by discretizing them. Finally, Section II introduces the hybrid scheme developed to efficiently solve this matrix system. Section III provides several numerical examples to demonstrate the computational benefits of the proposed HDG-BI solver. In Section IV, a short summary of the work is provided and several future research directions are briefly described.
\section{Formulation}
\subsection{Problem Description}\label{sec:prob}
Consider the electromagnetic scattering problem involving a dielectric object that resides in an unbounded background medium with permittivity $\varepsilon_0$ and permeability $\mu_0$ (Fig.~\ref{fig:oricomp}). The unbounded background medium is truncated into a finite computation domain that encloses the dielectric object. Let $\Omega$ and $\Gamma$ denote this computation domain and its boundary. In $\Omega$, the permittivity is given by $\varepsilon_0\epsilon_{\mathrm{r}}(\mathbf{r})$ and the permittivity is given by $\mu_0\mu_{\mathrm{r}}(\mathbf{r})$. Note that $\epsilon_{\mathrm{r}}(\mathbf{r})=1$ and $\mu_{\mathrm{r}}(\mathbf{r})=1$ in the background medium enclosed in $\Omega$ and $\epsilon_{\mathrm{r}}(\mathbf{r})\neq 1$ and $\mu_{\mathrm{r}}(\mathbf{r})\neq 1$ inside the scatterer. The speed of light in the background medium is given by $c_0 =1/\sqrt{\varepsilon_0 \mu_0}$.

The electric and magnetic fields incident on the object are represented by $\mathbf{E}^\mathrm{inc}(\mathbf{r})$ and $\mathbf{H}^\mathrm{inc}(\mathbf{r})$, respectively. It is assumed that the incident fields and all fields and currents generated as a result of this excitation are time-harmonic with time dependence $e^{j\omega t}$, where $t$ is the time and $\omega$ is the frequency of excitation. Let $\mathbf{E}^\mathrm{sca}(\mathbf{r})$ and $\mathbf{H}^\mathrm{sca}(\mathbf{r})$ denote the electric and magnetic fields scattered from the object, respectively. Then, one can express the total electric and magnetic fields as
$\mathbf{E}(\mathbf{r})=\mathbf{E}^\mathrm{inc}(\mathbf{r})+\mathbf{E}^\mathrm{sca}(\mathbf{r})$ and  $\mathbf{H}(\mathbf{r})=\mathbf{H}^\mathrm{inc}(\mathbf{r})+\mathbf{H}^\mathrm{sca}(\mathbf{r})$. On the computation domain boundary $\Gamma$, equivalent electric and magnetic currents are defined as $\mathbf{J}(\mathbf{r})=\hat{\mathbf{n}}_0(\mathbf{r})\times\mathbf{H}(\mathbf{r})$ and $\mathbf{M}(\mathbf{r})=-\hat{\mathbf{n}}_0(\mathbf{r})\times\mathbf{E}(\mathbf{r})$. Here, $\hat{\mathbf{n}}_0(\mathbf{r})$ is the outward-pointing unit normal vector on $\Gamma$. Note that the formulation presented in the rest of this section is derived for normalized electric fields $\mathbf{E}^\mathrm{inc}(\mathbf{r})$, $\mathbf{E}^\mathrm{sca}(\mathbf{r})$, and $\mathbf{E}(\mathbf{r})$, and the normalization factor is $\sqrt{{\epsilon_0}/{\mu_0}}$. The wavenumber in the background medium is given by $k_0=\omega\sqrt{\varepsilon_0 \mu_0}$.

The formulation presented in the rest of this section heavily uses two trace operators: (i) $\pi^{\tau}_{S}\{\mathbf{u}\}(\mathbf{r})=\hat{\mathbf{n}}(\mathbf{r}) \times\left.\mathbf{u}(\mathbf{r}) \times \hat{\mathbf{n}}(\mathbf{r})\right|_{S}$ that yields the tangential components of $\mathbf{u}(\mathbf{r})$ on surface $S$, and (ii) ${\pi^{ \times}_{S}\{\mathbf{u}\}(\mathbf{r})=\hat{\mathbf{n}}(\mathbf{r}) \times\left.\mathbf{u}(\mathbf{r})\right|_{S}}$ that yields the twisted tangential components of $\mathbf{u}(\mathbf{r})$ on $S$. Note that $\hat{\mathbf{n}}(\mathbf{r})$ is the outward-pointing unit normal vector on $S$.

Furthermore, to keep the formulation concise, the inner products used by the Galerkin scheme are not written explicitly. The notation and the definition of the inner products between two vectors $\mathbf{u}(\mathbf{r})$ and $\mathbf{v}(\mathbf{r})$ in volume $V$ and surface $S$ are given by
\begin{align}
\label{eq:9} \big(\mathbf{u}(\mathbf{r}), \mathbf{v}(\mathbf{r})\big)_{V}&=\int_{V}\! \mathbf{u}(\mathbf{r}) \cdot \mathbf{v}(\mathbf{r})\,dv\\
\label{eq:10}\big\langle\mathbf{u}(\mathbf{r}), \mathbf{v}(\mathbf{r})\big\rangle_{S}&=\int_{S}\! \mathbf{u}(\mathbf{r}) \cdot \mathbf{v}(\mathbf{r})\,ds
\end{align}
respectively.

In the rest of the formulation, the dependence of the variables and the operators on $\mathbf{r}$ is dropped for the sake of simplicity in the notation unless a new variable or an operator is introduced.

\subsection{Computation Domain Discretization}\label{sec:disc}
\begin{figure}[t!]
\centerline{\includegraphics[width=0.7\columnwidth,draft=false]{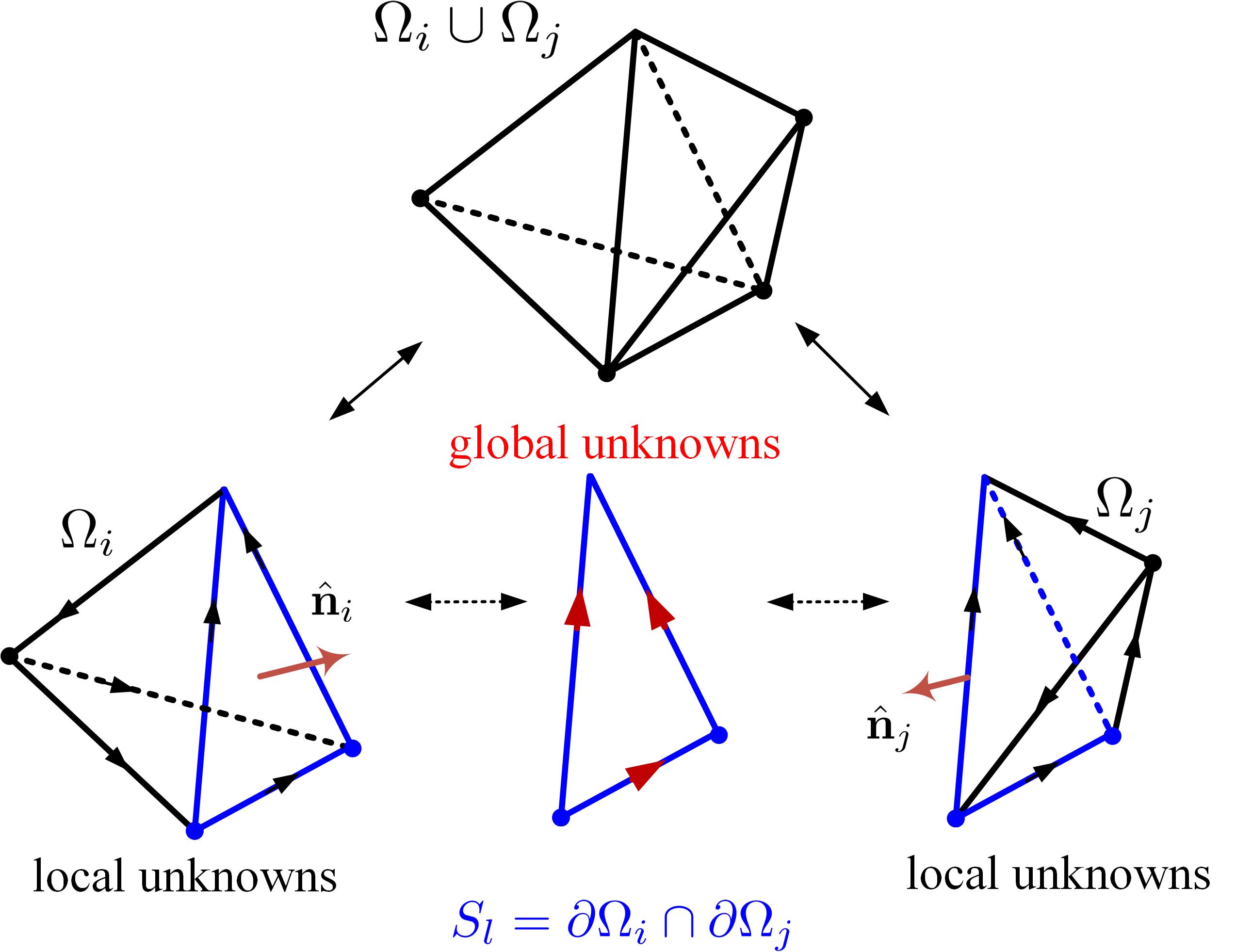}}
\caption{Description of the mesh supporting the local unknowns $\mathbf{E}_{\mathrm{h}}$ and $\mathbf{H}_{\mathrm{h}}$ and the global unknown $\boldsymbol{\Lambda}_{\mathrm{h}}$.}
\label{fig:disc}
\end{figure}

The computation domain $\Omega$ is discretized into a mesh of non-overlapping tetrahedrons represented by $\Omega_{\mathrm{h}}$: $\Omega \approx\Omega_\mathrm{h}= \bigcup_i \Omega_i$, where $\Omega_i$ is the $i$th tetrahedron. The boundary of tetrahedron $\Omega_i$, which consists of four triangular surfaces, is represented by $\partial \Omega_i$. $\mathbf{E}$ and $\mathbf{H}$ in $\Omega$ are approximated by $\mathbf{E}_{\mathrm{h}}$ and $\mathbf{H}_{\mathrm{h}}$ that are expanded on $\Omega_i$ of $\Omega_{\mathrm{h}}$.

The computation domain boundary $\Gamma$ is discretized into a mesh of non-overlapping triangular surfaces as represented by $\Gamma_{\mathrm{h}}$: $\Gamma_{\mathrm{h}}=\bigcup_i \Gamma_i$, where $\Gamma_i$ are the triangular surfaces of
$\Omega_{\mathrm{h}}$ that have all their three corners on $\Gamma$. $\mathbf{J}$ and $\mathbf{M}$ on $\Gamma$ are approximated by $\mathbf{J}_{\mathrm{h}}$ and $\mathbf{M}_{\mathrm{h}}$ that are expanded on pairs of $\Gamma_i$ of $\Gamma_{\mathrm{h}}$.

The traditional HDG method uses a global vector field, which is denoted by $\boldsymbol{\Lambda}_{\mathrm{h}}$, to ``connect'' local solutions $\mathbf{E}_{\mathrm{h}}$ and $\mathbf{H}_{\mathrm{h}}$ on $\Omega_i$ of $\Omega_{\mathrm{h}}$~\cite{nguyen2011hybridizable,li2014hybridizable}. The HDG-BI solver proposed in this work uses $\boldsymbol{\Lambda}_{\mathrm{h}}$ to also ``connect'' $\mathbf{E}_{\mathrm{h}}$ and $\mathbf{H}_{\mathrm{h}}$ in $\Omega_{\mathrm{h}}$ to $\mathbf{J}_{\mathrm{h}}$ and $\mathbf{M}_{\mathrm{h}}$ on $\Gamma_{\mathrm{h}}$. This global unknown $\boldsymbol{\Lambda}_{\mathrm{h}}$ is defined on the ``shared'' triangular surfaces of $\Omega_{\mathrm{h}}$ and the triangular surfaces of $\Gamma_{\mathrm{h}}$: $L_{\mathrm{h}}=S_{\mathrm{h}} \bigcup \Gamma_{\mathrm{h}}$, where $S_{\mathrm{h}}=\bigcup_l S_l$, $S_l$ is the triangular surface shared by two tetrahedrons $\Omega_i$ and $\Omega_j$, i.e., $S_l={\partial\Omega_i} \cap{\partial\Omega_j}$ (Fig.~\ref{fig:disc}), and $\Gamma_{\mathrm{h}}=\bigcup_i \Gamma_i$ (as already defined above).

\subsection{HDG-BI Formulation}\label{sec:hdgbi}
\subsubsection{HDG}
In computation domain $\Omega$, the electric field $\mathbf{E}$ and the magnetic field $\mathbf{H}$ satisfy the time-harmonic Maxwell's equations:
\begin{align}
\label{eq:1} j\omega \frac{\varepsilon_\mathrm{r} }{c_0}\mathbf{E}-\nabla \times \mathbf{H}&=0\\
\label{eq:2} j\omega \frac{\mu_\mathrm{r} }{c_0}\mathbf{H}+\nabla \times \mathbf{E}&=0.
\end{align}

\noindent Similar to the traditional DG schemes and FEM, HDG seeks $\mathbf{E}_\mathrm{h}$ and $\mathbf{H}_\mathrm{h}$ that are approximate solutions of~\eqref{eq:1} and~\eqref{eq:2} defined on mesh $\Omega_{\mathrm{h}}$ discretizing $\Omega$. This is achieved via weak Galerkin formulation of~\eqref{eq:1} and~\eqref{eq:2}. Let $\boldsymbol{e}$ and $\boldsymbol{h}$ represent the testing functions corresponding to $\mathbf{E}$ and $\mathbf{H}$ respectively. Then, in a given tetrahedron $\Omega_i$, one can express the weak form of~\eqref{eq:1} and~\eqref{eq:2} as
\begin{align}
\label{eq:11}\big({\boldsymbol{e}},  j\omega \frac{\varepsilon_\mathrm{r}}{c_0}\mathbf{E}_\mathrm{h}-\nabla \times \mathbf{H}_\mathrm{h} \big)_{\Omega_i}&=0 \\
\label{eq:12}\big({\boldsymbol{h}}, j\omega \frac{\mu_\mathrm{r}}{c_0}\mathbf{H}_\mathrm{h}+\nabla \times \mathbf{E}_\mathrm{h} \big)_{\Omega_i}&=0.
\end{align}
Using the mathematical identity for the divergence of the cross product of two vectors on the second terms of the inner products and applying the divergence theorem to the resulting expressions~\cite{Jin2015FEM}, one can convert~\eqref{eq:11} and~\eqref{eq:12} into
\begin{equation}
\label{eq:13}
\big( \boldsymbol{e}, j\omega \frac{{{\varepsilon }_{\mathrm{r}}}}{c_0}{{\mathbf{E}}_{\mathrm{h}}} \big)_{\Omega_i}-{\big( \nabla \times \boldsymbol{e}, {{\mathbf{H}}_{\mathrm{h}}} \big)_{\Omega_i}}
+{\big\langle  \pi^{\times}_{\partial \Omega_{i}} \!\{\boldsymbol{e}\}, \mathbf{H}_{\mathrm{h}}^{\mathrm{t}} \big\rangle_{\partial \Omega_i}}=0
\end{equation}
\begin{equation}
\label{eq:14}
\big( \boldsymbol{h}, j\omega \frac{{{\mu  }_{\mathrm{r}}}}{c_0}{{\mathbf{H}}_{\mathrm{h}}} \big)_{\Omega_i}+{\big( \nabla \times \boldsymbol{h}, {{\mathbf{E}}_{\mathrm{h}}} \big)_{\Omega_i}}
-{\big\langle \pi^{\times}_{\partial \Omega_{i}} \!\{\boldsymbol{h}\}, \mathbf{E}_{\mathrm{h}}^{\mathrm{t}} \big\rangle_{\partial \Omega_i}}=0.
\end{equation}
Here, $\mathbf{E}_\mathrm{h}^\mathrm{t}$ and $\mathbf{H}_\mathrm{h}^\mathrm{t}$ are the tangential components of $\mathbf{E}_\mathrm{h}$ and $\mathbf{H}_\mathrm{h}$ on ${\partial \Omega_i}$ and are expressed as $\mathbf{E}_\mathrm{h}^\mathrm{t}=\pi_{\partial\Omega_i}^\tau\!\{\mathbf{E}_{\mathrm{h}}\}$ and $\mathbf{H}_\mathrm{h}^\mathrm{t}=\pi_{\partial \Omega_i}^\tau\!\{\mathbf{H}_{\mathrm{h}}\}$, respectively.
To ``couple'' the local system of equations associated with tetrahedron $\Omega_i$ in~\eqref{eq:13} and~\eqref{eq:14} to the global system equations, numerical fluxes $\hat{\mathbf{H}}_\mathrm{h}^\mathrm{t}$ and $\hat{\mathbf{E}}_\mathrm{h}^\mathrm{t}$ are introduced as~\cite{nguyen2011hybridizable,li2014hybridizable}:
\begin{align}
\label{eq:15}
\hat{\mathbf{H}}_\mathrm{h}^\mathrm{t}&=\boldsymbol{\Lambda}_\mathrm{h}\\
\label{eq:16}\hat{\mathbf{E}}_\mathrm{h}^\mathrm{t}&=\mathbf{E}_\mathrm{h}^\mathrm{t}+\tau\pi_{\partial \Omega_i}^{\times}\!\{\boldsymbol{\Lambda}_\mathrm{h}-\mathbf{H}_\mathrm{h}^\mathrm{t}\}.
\end{align}
where $\tau$ is a local stabilization parameter and it is set to $1.0$ in the rest of formulation and the code that implements this formulation.
Unlike the traditional DG schemes, where the local fields $\mathbf{E}_\mathrm{h}$ and $\mathbf{H}_\mathrm{h}$ of a given tetrahedron are coupled to those of its neighboring tetrahedrons via numerical fluxes that rely on mean and jump of the field values~\cite{hesthaven2007nodal}, the numerical fluxes used by HDG as described in~\eqref{eq:15} and~\eqref{eq:16} couple the local fields $\mathbf{E}_\mathrm{h}$ and $\mathbf{H}_\mathrm{h}$ and the global unknown $\boldsymbol{\Lambda}_{\mathrm{h}}$.

Next, expressions of $\hat{\mathbf{E}}_\mathrm{h}^\mathrm{t}$ and $\hat{\mathbf{H}}_\mathrm{h}^\mathrm{t}$ in ~\eqref{eq:15} and~\eqref{eq:16} are used to replace $\mathbf{H}_\mathrm{h}^\mathrm{t}$ and $\mathbf{E}_\mathrm{h}^\mathrm{t}$ in~\eqref{eq:13} and~\eqref{eq:14}, respectively. This yields
\begin{equation}
\label{eq:17a}
\big(\boldsymbol{e}, j \omega \frac{\varepsilon_\mathrm{r} }{c_0} \mathbf{E}_\mathrm{h}\big)_{\Omega_i}-\big(\nabla \times \boldsymbol{e}, \mathbf{H}_\mathrm{h}\big)_{\Omega_i}+
\big\langle \pi^{\times}_{\partial \Omega_{i}} \!\{\boldsymbol{e}\}, \boldsymbol{\Lambda}_\mathrm{h}\big\rangle_{\partial {\Omega_i}}=0
\end{equation}
\begin{equation}
\label{eq:17b}
\begin{aligned}
\big(\boldsymbol{h}, j \omega \frac{\mu_\mathrm{r} }{c_0} \mathbf{H}_\mathrm{h}\big)_{\Omega_i}+\big(\boldsymbol{h}, \nabla \times \mathbf{E}_\mathrm{h}\big)_{\Omega_i}+
\big\langle \pi^{\times}_{\partial \Omega_{i}} \!\{\boldsymbol{h}\},  \pi^{\times}_{\partial \Omega_{i}}\!\{\mathbf{H}_\mathrm{h}-\boldsymbol{\Lambda}_\mathrm{h}\}\big\rangle_{\partial \Omega_i}=0.
\end{aligned}
\end{equation}
Note that the inner product $(\boldsymbol{h}, \nabla \times \mathbf{E}_\mathrm{h})_{\Omega_i}$ in~\eqref{eq:17b} is obtained after applying the divergence theorem to $\langle\pi_{\partial \Omega_i}^{\times}\!\{\boldsymbol{h}\}, \mathbf{E}_{\mathrm{h}}^{\mathrm{t}}\rangle_{\partial \Omega_i}$ and using the mathematical identity for the divergence of the cross product of two vectors on the resulting expression.

To ensure the continuity between local and global unknowns, one needs to enforce the field continuity condition on triangular surfaces of $\Omega_h$, namely $S_{\mathrm{h}}=\bigcup_l S_l$ and $\Gamma_{\mathrm{h}}=\bigcup_j \Gamma_j$. On a given shared/inner triangular surface $S
_l=\partial \Omega_i \cap \partial \Omega_j$, the continuity of the fields in $\Omega_i$ and $\Omega_j$ is enforced using the numerical flux~\cite{li2013hybridizable,li2014hybridizable}
\begin{equation}
\label{eq:18a}
\begin{aligned}
&\pi _{\partial {{\Omega}_{i}}}^{\times }\!\{\mathbf{E}_{\mathrm{h}} \}+\pi _{\partial \Omega_i}^{\tau }\!\{\mathbf{H}_{\mathrm{h}}-{{\boldsymbol{\Lambda }}_{\mathrm{h}}}\}+\\
&\pi _{\partial {{\Omega}_{j}}}^{\times }\!\{\mathbf{E}_{\mathrm{h}}\}+\pi _{\partial {{\Omega}_{j}}}^{\tau }\!\{\mathbf{H}_{\mathrm{h}}-{{\boldsymbol{\Lambda }}_{\mathrm{h}}} \}=0.
\end{aligned}
\end{equation}
On a given boundary triangular surface  $\Gamma_j = \partial \Omega_i \cap \Gamma_j$, the continuity of the fields in $\Omega_i$ across the computation domain boundary is enforced using the numerical flux
\begin{equation}
\begin{aligned}
&\pi _{\partial {{\Omega}_{i}}}^{\times }\!\{{{\mathbf{E}}_{\mathrm{h}}} \}+\pi _{\partial {{\Omega}_{i}}}^{\tau }\!\{{{\mathbf{H}}_{\mathrm{h}}}-{{\boldsymbol{\Lambda }}_{\mathrm{h}}} \}-\\ &\pi _{{{\Gamma}_j}}^{\tau }\!\{\mathbf{M}_{\mathrm{h}} \}-\pi _{{{\Gamma}_j}}^{\times }\!\{\mathbf{J}_{\mathrm{h}} \}-\pi _{{{\Gamma}_j}}^{\tau }\!\{{{\boldsymbol{\Lambda }}_{\mathrm{h}}} \}=0.\\
\end{aligned}
\label{eq:18b}
\end{equation}
Let $\boldsymbol{\eta}$ represent the testing function corresponding to the global unknown ${{\boldsymbol{\Lambda }}_{\mathrm{h}}}$, Then, one can express the weak form of~\eqref{eq:18a} and~\eqref{eq:18b} as
\begin{equation}
\label{eq:20}
\begin{aligned}
&\big\langle\boldsymbol{\eta}, \pi^{\times}_{\partial \Omega_{i}}\!\{\mathbf{E}_\mathrm{h}\}\big\rangle_{S_l}+\big\langle\boldsymbol{\eta},\pi^{\tau}_{\partial \Omega_{i}}\!\{\mathbf{H}_\mathrm{h}-\boldsymbol{\Lambda}_\mathrm{h}\}\big\rangle_{S_l}+\\
&\big\langle\boldsymbol{\eta}, \pi^{\times}_{\partial \Omega_{j}}\!\{\mathbf{E}_\mathrm{h}\}\big\rangle_{S_l}+
\big\langle\boldsymbol{\eta},\pi^{\tau}_{\partial \Omega_{j}}\!\{\mathbf{H}_\mathrm{h}-\boldsymbol{\Lambda}_\mathrm{h}\}\big\rangle_{S_l}
=0
\end{aligned}
\end{equation}
\begin{equation}
\label{eq:21}
\begin{aligned}
\big\langle\boldsymbol{\eta}, \pi^{\times}_{\partial \Omega_{i}}\!\{\mathbf{E}_\mathrm{h}\}\big\rangle_{\Gamma_j}+\big\langle\boldsymbol{\eta}, \pi^{\tau}_{\partial \Omega_{i}}\!\{\mathbf{H}_\mathrm{h}-\boldsymbol{\Lambda}_\mathrm{h}\}\big\rangle_{\Gamma_j}-& \\
\big\langle\boldsymbol{\eta}, \pi^{\tau}_{\Gamma_j}\{\mathbf{M}_\mathrm{h}\}+ \pi^{\times}_{\Gamma_j}\!\{\mathbf{J}_\mathrm{h}\}+ \pi^{\tau}_{\Omega_j}\!\{\boldsymbol{\Lambda}_\mathrm{h}\} \big\rangle_{\Gamma_j}=0
\end{aligned}
\end{equation}
By collecting the weak forms for all tetrahedrons $\Omega_i$ of $\Omega_{\mathrm{h}}$ and all triangular surfaces $S_l$ of $S_{\mathrm{h}}$ and $\Gamma_j$ of $\Gamma_{\mathrm{h}}$, one can obtain the part of the matrix system that represents the HDG component of the HDG-BI solver~\cite{nguyen2011hybridizable,li2014hybridizable}. This matrix system and the hybrid method used to efficiently solve it are described in Section~\ref{sec:matrix}.
\subsubsection{BI}
The formulation of the governing equations for the BI component of the proposed solver starts with the well-known relationship between the scattered fields $\mathbf{E}^{\mathrm{sca}}$ and $\mathbf{H}^{\mathrm{sca}}$ and the equivalent currents $\mathbf{J}$ and $\mathbf{M}$ that are introduced on the computation domain boundary $\Gamma$~\cite{Jin2012}:
\begin{align}
\label{eq:22a}\mathbf{E}^{\mathrm{sca}}(\mathbf{r}) &= \mathcal{L}_S\{\mathbf{J}\}(\mathbf{r}) -\mathcal{K}_S\{\mathbf{M}\}(\mathbf{r})\\
\label{eq:22b}\mathbf{H}^{\mathrm{sca}}(\mathbf{r}) & =\mathcal{K}_S\{\mathbf{J}\}(\mathbf{r}) +\mathcal{L}_S\{\mathbf{M}\}(\mathbf{r}).
\end{align}
Here, the integral operators $\mathcal{L}_S\{\mathbf{X}\}(\mathbf{r})$ and $\mathcal{K}_S\{\mathbf{X}\}(\mathbf{r})$ are given by
\begin{align*}
\mathcal{L}_S\{\mathbf{X}\}(\mathbf{r})&= -j k_{0} \int_{S}\!\left[\mathbf{I}+\frac{1}{k_{0}^{2}} \nabla \nabla \cdot\right] \mathbf{X}(\mathbf{r}^{\prime})G_{0}(\mathbf{r}, \mathbf{r}^{\prime}) \,ds^{\prime}\\
\mathcal{K}_S\{\mathbf{X}\}(\mathbf{r})&=\int_{S}\! \nabla G_{0}(\mathbf{r}, \mathbf{r}^{\prime}) \times \mathbf{X}(\mathbf{r}^{\prime}) \,ds^{\prime}
\end{align*}
where $G_{0}$ is theGreen's function of the unbounded medium with wavenumber $k_0$. Note that $\mathcal{K}_S\{\mathbf{X}\}= \mathbf{X}/2 \times \mathbf{\hat{n}}+\overline{\mathcal{K}}_S\{\mathbf{X}\}$ where $\overline{\mathcal{K}}_S$ is the principle value of ${\mathcal{K}}_S$.

Inserting~\eqref{eq:22a} and~\eqref{eq:22b} into the current-field relationships $\mathbf{M}=-\hat{\mathbf{n}}_0\times(\mathbf{E}^\mathrm{inc}+\mathbf{E}^\mathrm{sca})$ and $\mathbf{J}=\hat{\mathbf{n}}_0\times(\mathbf{H}^\mathrm{inc}+\mathbf{H}^\mathrm{sca})$ on $\Gamma$ yields the electric-field integral equation (EFIE) and the magnetic-field integral equation (MFIE), respectively~\cite{Jin2012,yla2005application,peng2012computations,sheng1998formulation}:
\begin{align}
\label{eq:24a}&\frac{1}{2} \pi^{\times}_{\Gamma}\{\mathbf{M}\}-\pi^{\tau}_{\Gamma}\{\mathcal{L}_{\Gamma}\{\mathbf{J}\}- \overline{\mathcal{K}}_{\Gamma}\{\mathbf{M}\}\}=\pi^{\tau}_{\Gamma}\{\mathbf{E}^\mathrm{i n c}\}\\
\label{eq:24b}-&\frac{1}{2} \pi^{\times}_{\Gamma} \{\mathbf{J}\}-\pi^{\tau}_{\Gamma} \{\mathcal{L}_{\Gamma}\{\mathbf{M} \}+ \overline{\mathcal{K}}_{\Gamma}\{\mathbf{J} \}\}=\pi^{\tau}_{\Gamma}\{ \mathbf{H}^\mathrm{i n c}\}.
\end{align}
To obtain a better-conditioned matrix upon discretization, linear combinations $\alpha \mathrm{EFIE}(19) + (1-\alpha) \hat{\mathbf{n}}_0\times\mathrm{MFIE} (20)$ and $(1-\alpha) \hat{\mathbf{n}}_0\times \mathrm{EFIE}(19)+\alpha\mathrm{MFIE}(20)$
are used to yield the electric current combined-field integral equation (JCFIE) and the magnetic current combined-field integral equation (MCFIE), respectively~\cite{yla2005application,peng2012computations,sheng1998formulation}:
\begin{equation}
\alpha \pi^{\times}_{\Gamma}\{\mathbf{M}\}+(1-\alpha) \mathbf{J}-\pi^{\times}_{\Gamma}\{\mathcal{C}_{\Gamma}^{(1-\alpha)}\{\mathbf{M} \}\}
-\mathcal{C}^{\alpha}_{\Gamma}\{\mathbf{J}\}=\alpha \pi^{\tau}_{\Gamma} \{\mathbf{E}^\mathrm{{i n c}}\} +(1-\alpha) \pi^{\times}_{\Gamma} \{\mathbf{H}^\mathrm{i n c}\}
\label{eq:25}
\end{equation}
\begin{equation}
(1-\alpha) \mathbf{M}-\alpha \pi^{\times}_{\Gamma}\{\mathbf{J}\}- \mathcal{C}_{\Gamma}^{\alpha}\{\mathbf{M} \}+
\pi^{\times}_{\Gamma}\{\mathcal{C}_{\Gamma}^{(1-\alpha)}\{\mathbf{J}\}\}=\alpha \pi^{\tau}_{\Gamma} \{\mathbf{H}^\mathrm{i n c}\}-(1-\alpha) \pi^{\times}_{\Gamma} \{\mathbf{E}^\mathrm{i n c}\}.
\label{eq:26}
\end{equation}
Here, the combined-field integral operator $\mathcal{C}^{\alpha}_S \{\mathbf{X}\}(\mathbf{r})$ is defined as~\cite{peng2012computations}
\begin{equation}
\mathcal{C}^{\alpha}_S \{\mathbf{X}\}(\mathbf{r}) =\alpha \pi^{\tau}_{S}\{\mathcal{L}_S\{\mathbf{X}\}(\mathbf{r})\}
+(1-\alpha) \pi^{\times}_{S}\{\overline{\mathcal{K}}_S \{\mathbf{X}\}(\mathbf{r})\}
\label{eq:27}
\end{equation}
and $\alpha$ is the weight that should be selected as $0 \leq \alpha \leq 1$. In the rest of formulation and the code that implements this formulation, $\alpha=0.5$. Accordingly $\mathcal{C}^{\alpha}_S\{\mathbf{X}\}$ is simplified as $\mathcal{C}_S\{\mathbf{X}\}$. JCFIE~\eqref{eq:25} and MCFIE~\eqref{eq:26} are approximated on $\Gamma_{\mathrm{h}}$ that discretizes $\Gamma$:
\begin{equation}
\frac{1}{2}\pi^{\times}_{\Gamma_{\mathrm{h}}}\!\{\mathbf{M}_{\mathrm{h}}\}+\frac{1}{2}\mathbf{J}_{\mathrm{h}}-\pi^{\times}_{\Gamma_{\mathrm{h}}}\{\mathcal{C}_{\Gamma_{\mathrm{h}}}\!\{\mathbf{M}_{\mathrm{h}} \}\}
-\mathcal{C}_{\Gamma_{\mathrm{h}}}\!\{\mathbf{J}_{\mathrm{h}}\}=\frac{1}{2}\pi^{\tau}_{\Gamma_{\mathrm{h}}}\!\{\mathbf{E}^\mathrm{{i n c}}\} +\frac{1}{2}\pi^{\times}_{\Gamma_{\mathrm{h}}}\!\{\mathbf{H}^\mathrm{i n c}\}
\label{eq:25a}
\end{equation}
\begin{equation}
\frac{1}{2}\mathbf{M}_{\mathrm{h}}-\frac{1}{2}\pi^{\times}_{\Gamma_{\mathrm{h}}}\!\{\mathbf{J}_{\mathrm{h}}\}- \mathcal{C}_{\Gamma_{\mathrm{h}}}\!\{\mathbf{M}_{\mathrm{h}} \}+
\pi^{\times}_{\Gamma_{\mathrm{h}}}\!\{\mathcal{C}_{\Gamma_{\mathrm{h}}}\{\mathbf{J}_{\mathrm{h}}\}\}=\frac{1}{2}\pi^{\tau}_{\Gamma_{\mathrm{h}}}\!\{\mathbf{H}^\mathrm{i n c}\}-\frac{1}{2}\pi^{\times}_{\Gamma_{\mathrm{h}}}\!\{\mathbf{E}^\mathrm{i n c}\}.
\label{eq:26a}
\end{equation}
The continuity of the fields on $\Gamma_{\mathrm{h}}$ is enforced using the numerical flux
\begin{equation}
\label{eq:28}\mathbf{J}_{\mathrm{h}}-\pi^{\times}_{\Gamma_{\mathrm{h}}}\!\{\boldsymbol{\Lambda}_{\mathrm{h}} \} =0.
\end{equation}
The governing BI equations are obtained via two linear combinations: $\mathrm{JCFIE}(\ref{eq:25a})+\frac{1}{2}(\ref{eq:28})$ and $\mathrm{MCFIE}(\ref{eq:26a})+\frac{1}{2}\pi_{\Gamma_{\mathrm{h}}}^{\times}\{(\ref{eq:28})\}$. This yields
\begin{equation}
\label{eq:30}
\mathbf{J}_{\mathrm{h}}-\mathcal{C}_{\Gamma_{\mathrm{h}}}\!\{\mathbf{J}_{\mathrm{h}}\}+\frac{1}{2} \pi_{\Gamma_{\mathrm{h}}}^{\times}\{\mathbf{M}_{\mathrm{h}}\}-\pi^{\times}_{\Gamma_{\mathrm{h}}}\!\{\mathcal{C}_{\Gamma_{\mathrm{h}}}\!\{\mathbf{M}_{\mathrm{h}}\}\}
-\frac{1}{2}\pi^{\times}_{\Gamma_{\mathrm{h}}}\!\{\boldsymbol{\Lambda}_{\mathrm{h}}\}=\frac{1}{2}\pi^{\tau}_{\Gamma_{\mathrm{h}}}\!\{\mathbf{E}^\mathrm{i n c}\} +\frac{1}{2}\pi^{\times}_{\Gamma_{\mathrm{h}}}\{\mathbf{H}^\mathrm{i n c}\}
\end{equation}
\begin{equation}
\label{eq:31}
\frac{1}{2} \mathbf{M}_{\mathrm{h}}- \mathcal{C}_{\Gamma_{\mathrm{h}}} \!\{\mathbf{M}_{\mathrm{h}}  \}
+\pi^{\times}_{\Gamma_{\mathrm{h}}} \!\{\mathcal{C}_{\Gamma_{\mathrm{h}}} \!\{\mathbf{J}_{\mathrm{h}}  \}\}+
\frac{1}{2}\pi^{\tau}_{\Gamma_{\mathrm{h}}}\!\{ \boldsymbol{\Lambda}_{\mathrm{h}}\}
=\frac{1}{2} \pi^{\tau}_{\Gamma_{\mathrm{h}}} \!\{\mathbf{H}^\mathrm{i n c}\} -\frac{1}{2} \pi^{\times}_{\Gamma_{\mathrm{h}}} \!\{\mathbf{E}^\mathrm{inc}\}.
\end{equation}
Let $\boldsymbol{j}$ and $\boldsymbol{m}$ represent the testing functions corresponding to $\mathbf{J}$ and $\mathbf{M}$, respectively. $\boldsymbol{j}$ and $\boldsymbol{m}$ are the well-known Rao-Wilton-Glisson (RWG) basis functions~\cite{rao1982electromagnetic} that are defined on pairs of $\Gamma_i$. Let each of these pairs be represented by $T_j$. Then, one can express the weak forms of~\eqref{eq:30} and~\eqref{eq:31} as
\begin{equation}
\begin{aligned}
&\big\langle \mathbf{j}, \mathbf{J}_\mathrm{h}-\mathcal{C}_{\Gamma_\mathrm{h}}\!\{\mathbf{J}_\mathrm{h} \}\big\rangle_{T_j}
-\big\langle \mathbf{j},\frac{1}{2}\pi^{\times}_{T_j}\!\{\boldsymbol{\Lambda}_\mathrm{h}\}\big\rangle_{T_j}
 +\big\langle\mathbf{j}
,\frac{1}{2}\pi^{\times}_{\Gamma_{\mathrm{h}}}\!\{\mathbf{M}_{\mathrm{h}}\}-\pi^{\times}_{T_j}\!\{\mathcal{C}_{\Gamma_\mathrm{h}}\!\{\mathbf{M}_\mathrm{h}\}\}\big\rangle_{T_j}\\
&=\big\langle \mathbf{j},\frac{1}{2} \pi^{\tau}_{T_j}\!\{\mathbf{E}^\mathrm{i n c}\} +\frac{1}{2} \pi^{\times}_{T_j}\!\{\mathbf{H}^\mathrm{i n c} \}\big\rangle_{T_j}
\end{aligned}
\label{eq:32}
\end{equation}
\begin{equation}
\begin{aligned}
&\big\langle \mathbf{m},\frac{1}{2} \mathbf{M}_\mathrm{h}- \mathcal{C}_{\Gamma_\mathrm{h}}\!\{\mathbf{M}_\mathrm{h} \}\big\rangle_{T_j}
+\big\langle \mathbf{m},\pi^{\times}_{T_j} \!\{\mathcal{C}_{\Gamma_\mathrm{h}} \!\{\mathbf{J}_\mathrm{h} \}\}\big\rangle_{T_j}
+\big\langle \mathbf{m},\frac{1}{2}\pi^{\tau}_{T_j}\!\{ \boldsymbol{\Lambda}_\mathrm{h}\}\big\rangle_{T_j} \\
&=\big\langle \mathbf{m},\frac{1}{2} \pi^{\tau}_{T_j} \!\{\mathbf{H}^\mathrm{i n c}\} -\frac{1}{2} \pi^{\times}_{T_j} \!\{\mathbf{E}^\mathrm{i n c}\}\big\rangle_{T_j}.
\end{aligned}
\label{eq:33}
\end{equation}
By collecting the weak forms for all pairs of triangular surfaces $T_j$ of $\Gamma_{\mathrm{h}}$, one can obtain the part of the matrix system that represents the BI component of the HDG-BI solver. This matrix system and the hybrid method used to efficiently solve it are described in Section~\ref{sec:matrix}.

\subsection{Matrix System}\label{sec:matrix}
On $\Omega_{\mathrm{h}}$, the local unknowns $\mathbf{E}_{\mathrm{h}}$ and $\mathbf{H}_{\mathrm{h}}$ are expanded as
\begin{align}
\label{eq:exp1} \mathbf{E}_\mathrm{h}&=\sum_i \bar{E}_i \boldsymbol{e}_i\\
\label{eq:exp2} \mathbf{H}_\mathrm{h}&=\sum_i \bar{H}_i \boldsymbol{h}_i
\end{align}
where $\boldsymbol{e}$ and $\boldsymbol{h}$ are the 3D zeroth-order vector edge basis functions~\cite{Jin2015FEM} and $\bar{E}$ and $\bar{H}$ are the vectors storing the corresponding expansion coefficients. Similarly, on $S_{\mathrm{h}}$ and $\Gamma_{\mathrm{h}}$, the global unknown $\boldsymbol{\Lambda}_\mathrm{h}$ is expanded as
\begin{equation}
\label{eq:exp3} \boldsymbol{\Lambda}_\mathrm{h}=\sum_i \bar{\mit{\Lambda}}^\mathrm{S}_i \boldsymbol{\eta}_i+\sum_j \bar{\mit{\Lambda}}^\mathrm{\Gamma}_j \boldsymbol{\eta}_j
\end{equation}
where $\boldsymbol{\eta}$ is the 2D zeroth-order vector edge basis function~\cite{Jin2015FEM} and $\bar{\mit{\Lambda}}^\mathrm{S}$ and $\bar{\mit{\Lambda}}^\mathrm{\Gamma}$ are the vectors storing the coefficients of the expansions on $S_{\mathrm{h}}$ and $\Gamma_{\mathrm{h}}$, respectively.

On $\Gamma_{\mathrm{h}}$, $\mathbf{J}_{\mathrm{h}}$ and $\mathbf{M}_{\mathrm{h}}$ are expanded as
\begin{align}
\label{eq:exp4} \mathbf{J}_{\mathrm{h}}&=\sum_i \bar{J}_i \boldsymbol{j}_i\\
\label{eq:exp5} \mathbf{M}_{\mathrm{h}}&=\sum_i \bar{M}_i \boldsymbol{m}_i
\end{align}
where $\boldsymbol{j}$ and $\boldsymbol{m}$ are the well-known RWG basis functions~\cite{rao1982electromagnetic} as mentioned earlier.


Inserting the expansions in~\eqref{eq:exp1}-\eqref{eq:exp5} into the weak forms~\eqref{eq:17a},~\eqref{eq:17b},~\eqref{eq:20}, and~\eqref{eq:21}, and collecting the resulting equations for all tetrahedrons $\Omega_i$ of $\Omega_{\mathrm{h}}$ and all triangular surfaces $S_l$ of $S_{\mathrm{h}}$ and $\Gamma_j$ of $\Gamma_{\mathrm{h}}$, one can obtain the part of the matrix system that represents the HDG component of the HDG-BI solver:
\begin{equation}
\bar{\bar{A}}\begin{bmatrix}
\bar{E} \\
\bar{H}
\end{bmatrix}+\bar{\bar{F}} \begin{bmatrix} \bar{\mit{\Lambda}^\mathrm{S}} \\ \bar{\mit{\Lambda}^\mathrm{\Gamma}} \end{bmatrix}=\bar{0}
\label{eq:34}
\end{equation}
\begin{equation}
\bar{\bar{B}}\begin{bmatrix}
\bar{E} \\
\bar{H}
\end{bmatrix}+\bar{\bar{L}} \begin{bmatrix}\bar{\mit{\Lambda}^\mathrm{S}} \\ \bar{\mit{\Lambda}^\mathrm{\Gamma} } \end{bmatrix}=-\begin{bmatrix}
\bar{\bar{0}} & \bar{\bar{0}}\\
\bar{\bar{D}}^\mathrm{\Lambda J} & \bar{\bar{D}}^\mathrm{\Lambda M}
\end{bmatrix}\begin{bmatrix}
\bar{J} \\
\bar{M}
\end{bmatrix}.
\label{eq:35}
\end{equation}
Here, the entries of the matrices $\bar{\bar{A}}$, $\bar{\bar{F}}$, $\bar{\bar{B}}$, $\bar{\bar{L}}$, $\bar{\bar{D}}^\mathrm{\Lambda J}$, and $\bar{\bar{D}}^\mathrm{\Lambda M}$ are given by
\begin{equation}
\!\!\!\bar{\bar{A}}_{mn}=\begin{bmatrix}
	\big( \boldsymbol{e}_m,j\omega \frac{\varepsilon_\mathrm{r} }{c_0}\boldsymbol{e}_n \big) _{\Omega_i}\!\!&\!\!		-\big( \nabla \times \boldsymbol{e}_m,\boldsymbol{h}_n \big) _{\Omega_i}\\
	\big( \boldsymbol{h}_m,\nabla \times \boldsymbol{e}_n \big) _{\Omega_i}\!\!&\!\! \left\{ \begin{aligned}
	&\big( \boldsymbol{h}_m,j\omega \frac{\mu_\mathrm{r} }{c_0}\boldsymbol{h}_n \big) _{\Omega_i}\\
	+&\big\langle \boldsymbol{h}_m,\pi^{\tau}_{\partial {\Omega_i}}\!\{\boldsymbol{h}_n\}\big \rangle _{\partial {\Omega_i}}\\
\end{aligned} \right\}\\
\end{bmatrix}
\label{eq:36}
\end{equation}
\begin{equation}
\bar{\bar{F}}_{mn}=
\begin{bmatrix}
\big\langle \pi^{\times}_{\partial {\Omega_i}}\!\{\boldsymbol{e}_m\}, \boldsymbol{\eta}_n\big\rangle_{\partial {\Omega_i}} \\
-\big\langle \pi^{\times}_{\partial {\Omega_i}} \!\{\boldsymbol{h}_m\}, \pi^{\times}_{\partial {\Omega_i}}\!\{\boldsymbol{\eta}_n\}\big\rangle_{\partial \Omega_i}
\end{bmatrix}
\label{eq:39b}
\end{equation}
\begin{equation}
\bar{\bar{B}}_{mn}=\begin{bmatrix}\big\langle\boldsymbol{\eta}_m, \pi^{\times}_{\partial {\Omega_i}}\!\{\boldsymbol{e}_n\}\big\rangle_{S_l}  \quad \big\langle\boldsymbol{\eta}_m, \pi^{\tau}_{\partial {\Omega_i}}\!\{\boldsymbol{h}_n\}\big\rangle_{S_l}\end{bmatrix}
\label{eq:38}
\end{equation}
\begin{equation}
\bar{\bar{L}}_{mn}=-\begin{bmatrix}2\big\langle\boldsymbol{\eta}_m, \pi^{\tau}_{\partial {\Omega_i}}\!\{\boldsymbol{\eta}_n\}\big\rangle_{S_l} \end{bmatrix}
\label{eq:39}
\end{equation}
\begin{equation}
\label{eq:40a} \bar{\bar{D}}^\mathrm{\Lambda J}_{mn}=-\begin{bmatrix} \big\langle\boldsymbol{\eta}_m,\pi^{\times}_{T_n}\!\{\mathbf{j}_n\}\big\rangle_{S_l}\end{bmatrix}
\end{equation}
\begin{equation}
\label{eq:40b} \bar{\bar{D}}^\mathrm{\Lambda M}_{mn}=-\begin{bmatrix}\big\langle\boldsymbol{\eta}_m, \pi^{\tau}_{T_n}\!\{\mathbf{m}_n\}\big\rangle_{S_l}\end{bmatrix}.
\end{equation}
To decrease the computational cost, the HDG scheme relies on reducing the size of the matrix system it solves. This is done by inverting~\eqref{eq:34} for $\begin{bmatrix} {\bar{E}}\; {\bar{H}} \end{bmatrix}^T$ and inserting the resulting expression into~\eqref{eq:35}. This operation yields:
\begin{equation}
\left(\bar{\bar{L}}-\bar{\bar{B}} \bar{\bar{A}}^{-1} \bar{\bar{F}}\right)\begin{bmatrix} \bar{\mit{\Lambda}^\mathrm{S}} \\ \bar{\mit{\Lambda}^\mathrm{\Gamma}} \end{bmatrix}+\begin{bmatrix}
\bar{\bar{0}} & \bar{\bar{0}}\\
\bar{\bar{D}}^\mathrm{\Lambda J} & \bar{\bar{D}}^\mathrm{\Lambda M}
\end{bmatrix}\begin{bmatrix}
\bar{J} \\
\bar{M}
\end{bmatrix}=\bar{0}.
\label{eq:41}
\end{equation}
Here, the dimension of the matrix $\bar{\bar{L}}-\bar{\bar{B}} \bar{\bar{A}}^{-1} \bar{\bar{F}}$ is equal to the number of degrees of freedom in the expansion of the global unknown $\boldsymbol{\Lambda}_\mathrm{h}$ as done in~\eqref{eq:exp3}. Let this number be represented by $N_{\mathrm{HDG}}$, and let the total number of degrees of freedom in the expansions of $\mathbf{E}_\mathrm{h}$ and $\mathbf{H}_\mathrm{h}$ as done in~\eqref{eq:exp1} and~\eqref{eq:exp2} be represented by $N_{\mathrm{DG}}$. Since $N_{\mathrm{HDG}} < N_{\mathrm{DG}}$, the computational cost of the HDG scheme is significantly smaller than that of the traditional DG schemes~\cite{nguyen2011hybridizable,li2014hybridizable}.


Inserting the expansions in~\eqref{eq:exp3}-\eqref{eq:exp5} into the weak forms~\eqref{eq:32} and~\eqref{eq:33} and collecting the resulting equations for all pairs of triangular surfaces $T_m$ of $\Gamma_{\mathrm{h}}$ yield the part of the matrix system that represents the BI component of the HDG-BI solver. Combining this system with the one for the HDG component in~\eqref{eq:41} yields the final matrix system of the HDG-BI solver as
\begin{equation}
\begin{aligned}
\renewcommand*{\arraystretch}{1.35}
\left[ \begin{array}{c|c}
\!\! \bar{\bar{L}}-\bar{\bar{B}}\bar{\bar{A}}^{-1} \bar{\bar{F}} \!\! &\!\! \begin{array}{cc}
\bar{\bar{0}} & \bar{\bar{0}}\\
\bar{\bar{D}}^\mathrm{\Lambda J} & \bar{\bar{D}}^\mathrm{\Lambda M}
\end{array} \!\!\!\! \\
\hline
\!\! \begin{array}{cc}
\bar{\bar{0}} &\bar{\bar{D}}^\mathrm{J \Lambda}\\
 \bar{\bar{0}}  &\bar{\bar{D}}^\mathrm{M \Lambda}
\end{array}\!\!\!\! & \!\! \begin{array}{cc}
\bar{\bar{C}}^\mathrm{JJ} & \bar{\bar{C}}^\mathrm{JM}\\
\bar{\bar{C}}^\mathrm{MJ} & \bar{\bar{C}}^\mathrm{MM} \\
\end{array} \!\!\!\! \\
\end{array} \right] \! \!
\left[\begin{array}{c}
\!\!\!\! \begin{array}{c}
\bar{\mit{\Lambda}^\mathrm{S}} \\
\bar{\mit{\Lambda}^\mathrm{\Gamma}}
\end{array}\!\!\!\! \\
\hline
\!\!\!\! \begin{array}{c}
\bar{J} \\
\bar{M}
\end{array}\!\!\!\!
\end{array}\right]=
\left[\begin{array}{c}
\!\!\!\! \begin{array}{c}
{\bar{0}} \\
{\bar{0}}
\end{array}\!\!\!\! \\
\hline
\!\!\!\! \begin{array}{c}
\bar{b}^\mathrm{J} \\
\bar{b}^\mathrm{M}
\end{array}\!\!\!\!
\end{array}\right].
\end{aligned}
\label{eq:42}
\end{equation}
Here, the entries of the matrices $\bar{\bar{D}}^\mathrm{J \Lambda}$, $\bar{\bar{D}}^\mathrm{M \Lambda}$, $\bar{\bar{C}}^\mathrm{J J }$, $\bar{\bar{C}}^\mathrm{J M }$, $\bar{\bar{C}}^\mathrm{M J }$, and $\bar{\bar{C}}^\mathrm{M M}$ and the entries of the right-hand side vectors ${\bar{b}}^\mathrm{J}$ and ${\bar{b}}^\mathrm{M}$ are given by
\begin{equation}
\bar{\bar{D}}^\mathrm{J \Lambda}_{mn}=-\big\langle\boldsymbol{j}_{m}, \frac{1}{2} \pi^{\times}_{T_m}\!\{\boldsymbol{\eta}_n\}\big\rangle_{T_m}
\label{eq:47}\end{equation}
\begin{equation}
\bar{\bar{D}}^\mathrm{M \Lambda}_{mn}=\big\langle\boldsymbol{m}_{m}, \frac{1}{2} \pi^{\tau}_{T_m}\!\{\boldsymbol{\eta}_n\}\big\rangle_{T_m}
\label{eq:48}\end{equation}
\begin{equation}
\bar{\bar{C}}_{mn}^\mathrm{JJ}=\big\langle\boldsymbol{j}_m, \boldsymbol{j}_n-\mathcal{C}_{T_n}\!\{\boldsymbol{j}_n\} \big\rangle_{{T_m}}
\label{eq:43}\end{equation}
\begin{equation}
\bar{\bar{C}}_{mn}^\mathrm{JM}=\big\langle\boldsymbol{j}_m,  \frac{1}{2}\pi^{\times}_{ T_n}\!\{\boldsymbol{m}_n\}-\pi^{\times}_{T_n}\!\{\mathcal{C}_{T_n}\!\{\boldsymbol{m}_n\}\} \big\rangle_{{T_m}}
\label{eq:44}\end{equation}
\begin{equation}
\bar{\bar{C}}_{mn}^\mathrm{MJ}=\big\langle\boldsymbol{m}_{m}, \pi^{\times}_{T_m}\!\{\mathcal{C}_{T_n}\{\boldsymbol{j}_n\}\}\big\rangle_{T_m}
\label{eq:45}\end{equation}
\begin{equation}
\bar{\bar{C}}_{mn}^\mathrm{MM}=\big\langle\boldsymbol{m}_{m},\frac{1}{2}\boldsymbol{m}_n-\mathcal{C}_{T_n}\!\{\boldsymbol{m}_n \}\big\rangle_{{T_m}}
\label{eq:46}
\end{equation}
\begin{equation}
\bar{b}_m^{\mathrm{J}}=\big\langle\boldsymbol{j}_m, \frac{1}{2}\pi^{\tau}_{T_m}\!\{\mathbf{E}^\mathrm{i n c}\}+\frac{1}{2}\pi^{\times}_{T_m}\!\{\mathbf{H}^\mathrm{i n c}\}\big\rangle_{T_m}
\label{eq:49}\end{equation}
\begin{equation}
\bar{b}_m^\mathrm{M}=\big\langle\boldsymbol{m}_m, \frac{1}{2} \pi^{\tau}_{T_m}\!\{\mathbf{H}^\mathrm{i n c}\}-\frac{1}{2} \pi^{\times}_{T_m}\!\{\mathbf{E}^\mathrm{inc}\} \big\rangle_{T_m}.
\label{eq:50}
\end{equation}
The dimension of the matrix system~\eqref{eq:42} is $N_{\mathrm{HDG}}+N_{\mathrm{BI}}$, where $N_{\mathrm{HDG}}$ is already defined above and $N_{\mathrm{BI}}$ is the total number of degrees of freedom in the expansions of $\mathbf{J}_\mathrm{h}$ and $\mathbf{M}_\mathrm{h}$ as done in~\eqref{eq:exp4} and~\eqref{eq:exp5}. This matrix system is solved for the unknown vectors $\bar{\mit{\Lambda}^\mathrm{S}}$, $\bar{\mit{\Lambda}^\mathrm{\Gamma}}$, $\bar{J}$, and $\bar{M}$ using the scheme described in Section~\ref{sec:solver}.

\begin{table}[t!]
\centering
\def\arraystretch{1.1}%
\caption{$N_{\mathrm{HDG}}$, $N_{\mathrm{BI}}$, and $N_{\mathrm{DG}}$ of the discretizations used in the electromagnetic simulation of scattering from the coated sphere. }
\label{t:5-0}
  \begin{tabular}{ c | c | c | c }
  Average edge length  & $N_{\mathrm{HDG}}$ & $N_{\mathrm{BI}}$ & $N_{\mathrm{DG}}$\\ \hline
  $0.1\lambda_0$ & $9\,351$  & $1\,596$ & $16\,392$  \\ \hline
  $0.075\lambda_0$ & $29\,757$  & $2\,592$ & $55\,356$  \\ \hline
  $0.05\lambda_0$ & $101\,343$  & $9\,282$ & $193\,404$ \\ \hline
  $0.03\lambda_0$ & $508\,638$ & $16\,290$ & $992\,100$
  \end{tabular}
\end{table}
\begin{figure}[t!]
\centerline{\includegraphics[width=0.9\columnwidth,draft=false]{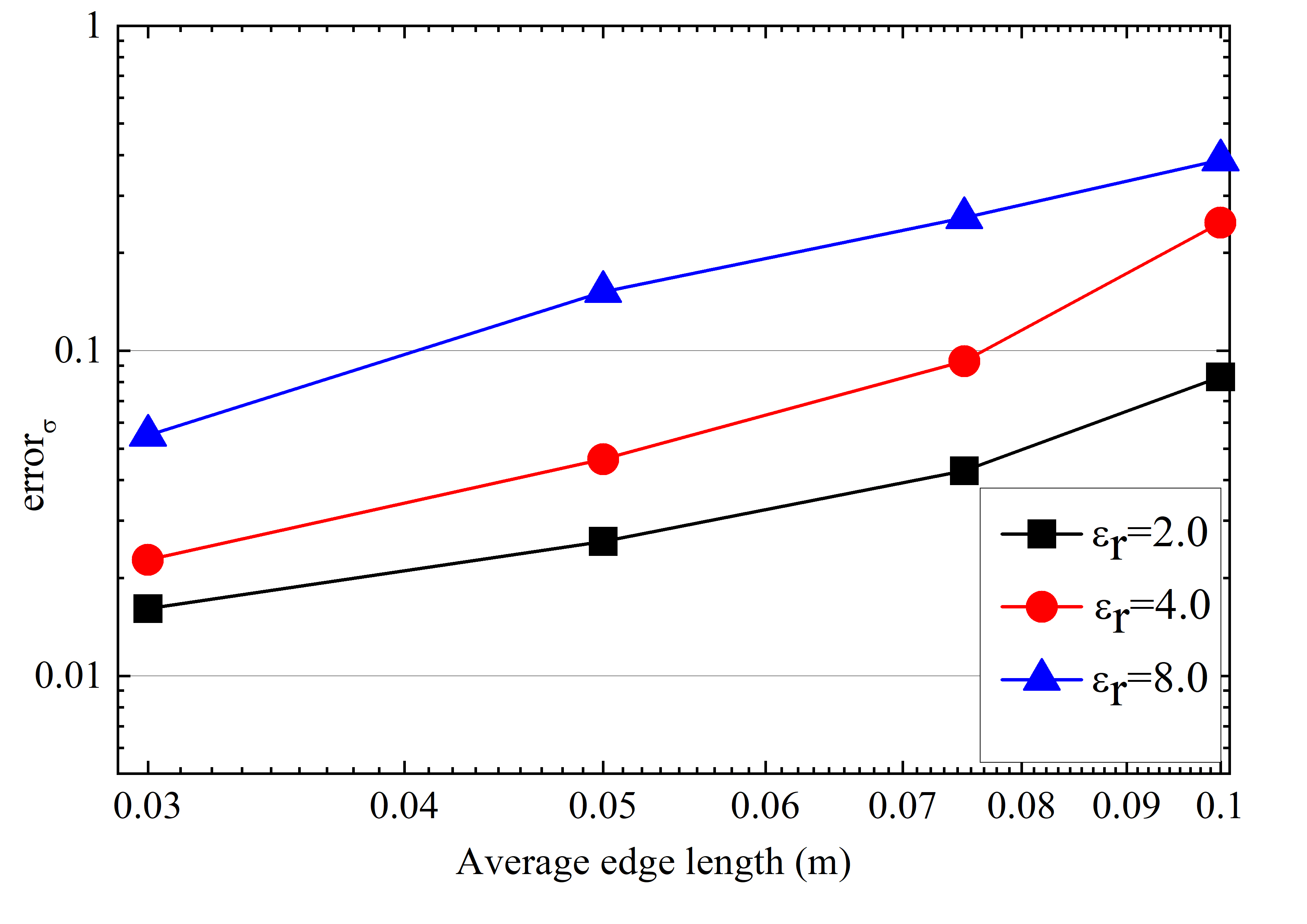}}
\caption{$L_2$-norm of the relative error in RCS $error_{\sigma}$ [computed using~\eqref{eq:5-25}] for different values of $\varepsilon_{\mathrm{r}}$ versus average edge length of discretization.}
\label{fig:rmseex1}
\end{figure}

\subsection{Hybrid Solver}\label{sec:solver}
The matrix system~\eqref{eq:42} can be re-written in a more compact form as:
\begin{equation}
\begin{bmatrix}
\bar{\bar{Q}} & \bar{\bar{D}}^\mathrm{\Lambda X} \\
\bar{\bar{D}}^\mathrm{X \Lambda} & \bar{\bar{C}}
\end{bmatrix}\begin{bmatrix}
\bar{\mit{\Lambda}} \\
\bar{X}
\end{bmatrix}=\begin{bmatrix}
0 \\
\bar{b}
\end{bmatrix}
\label{eq:51}
\end{equation}
where matrices $\bar{\bar{Q}}$, $\bar{\bar{D}}^{\mathrm{\Lambda X}}$, $\bar{\bar{D}}^{\mathrm{X \Lambda}}$, and $\bar{\bar{C}}$ represent the four blocks of the matrix in~\eqref{eq:42} and the vectors $\bar{\mit{\Lambda}}$, $\bar{X}$, and $\bar{b}$ represent the corresponding parts of the unknown and the right-hand side vectors.

$\bar{\bar{L}}$, $\bar{\bar{B}}$, and $\bar{\bar{F}}$ are sparse matrices and $\bar{\bar{A}}$ and $\bar{\bar{A}}^{-1}$ are block-diagonal matrices. Therefore, $\bar{\bar{Q}}$ is a sparse matrix. $\bar{\bar{D}}^{\mathrm{\Lambda X}}$ and $\bar{\bar{D}}^{\mathrm{X \Lambda}}$ are also sparse matrices since their blocks $\bar{\bar{D}}^{\mathrm{J \Lambda}}$, $\bar{\bar{D}}^{\mathrm{M \Lambda}}$, $\bar{\bar{D}}^{\mathrm{\Lambda J}}$, and $\bar{\bar{D}}^{\mathrm{\Lambda M}}$ are all sparse matrices. $\bar{\bar{C}}$ is a full matrix since its blocks $\bar{\bar{C}}^{\mathrm{JJ}}$, $\bar{\bar{C}}^{\mathrm{JM}}$, $\bar{\bar{C}}^{\mathrm{MJ}}$, and $\bar{\bar{C}}^{\mathrm{MJ}}$ are all full matrices.

Ideally, the matrix system~\eqref{eq:51} could be solved using a Krylov subspace-based iterative method assuming that the matrix-vector product associated with $\bar{\bar{C}}$ is accelerated using MLFMA~\cite{rokhlin1990rapid,lu1994multilevel,song1997multilevel,sheng1998solution,fostier2008,ergul2008,michiels2015,yucel2018,abduljabbar2019}. However, $\bar{\bar{Q}}$ is not well-conditioned~\cite{chen2020hybridizable} and as a result this iterative solution converges very slowly. Indeed, this is the reason why the traditional HDG schemes (in frequency domain) almost always rely on a direct (but sparse) matrix solver~\cite{li2013hybridizable,li2014hybridizable}. However, for the HDG-BI scheme developed in this work, using only a direct solver on~\eqref{eq:51} would be computationally expensive since $\bar{\bar{C}}$, which represents the BI component, is a full matrix.

To this end, in this work, a ``hybrid'' scheme is developed to efficiently solve the matrix system~\eqref{eq:51}. The first row of~\eqref{eq:51} is inverted for $\bar{\mit{\Lambda}}$ and the resulting expression is inserted into the second row to yield:
\begin{equation}
\left(-\bar{\bar{D}}^\mathrm{X \Lambda} \bar{\bar{Q}}^{-1} \bar{\bar{D}}^\mathrm{\Lambda X}+\bar{\bar{C}}\right)\bar{X}=\bar{b}
\label{eq:52}\end{equation}
This ``reduced'' matrix system of dimension $N_{\mathrm{BI}}$ is solved using the hybrid scheme as described next step by step:
\begin{enumerate}
\item Apply LU decomposition to the sparse matrix $\bar{\bar{Q}}$ as $\bar{\bar{Q}}=\bar{\bar{L}}\bar{\bar{U}}$ and store matrices $\bar{\bar{L}}$ and $\bar{\bar{U}}$.
\item Start the iterations of a Krylov subspace-based iterative scheme to solve~\eqref{eq:52}. The matrix-vector multiplication subroutine of this iterative scheme is implemented as described by steps (a)-(c) below. Let $\bar{x}_0$ be the guess vector of this matrix-vector multiplication.
\begin{enumerate}
\item Apply MLFMA~\cite{rokhlin1990rapid,lu1994multilevel,song1997multilevel,sheng1998solution,fostier2008,ergul2008,michiels2015,yucel2018,abduljabbar2019} to accelerate the matrix-vector product $\bar{y}_\mathrm{d}= \bar{\bar{C}}\bar{x}_0$.
\item Compute the matrix-vector product $\bar{x}_1=\bar{\bar{D}}^\mathrm{\Lambda X}\bar{x}_0$.
\item Compute the matrix-vector product $\bar{x}_2=\bar{\bar{Q}}^{-1}\bar{x}_1=\bar{\bar{U}}^{-1}\bar{\bar{L}}^{-1}\bar{x}_1$ in two steps: i) Solve $\bar{\bar{L}}\bar{v}=\bar{x}_{1}$ for $\bar{v}$ via forward substitution and ii) solve $\bar{\bar{U}}\bar{x}_{2}=\bar{v}$ for $\bar{x}_2$ via backward substitution.
\item Compute the matrix-vector product $\bar{y}_\mathrm{s}=-\bar{\bar{D}}^\mathrm{X \Lambda}\bar{x}_{2}$.
\item Compute the output of the matrix-vector multiplication subroutine as $\bar{y} = \bar{y}_\mathrm{d}+\bar{y}_\mathrm{s}$.
\end{enumerate}
\item Continue the iterations until the relative residual error reaches the desired level.
\end{enumerate}
The iterative solver used above is the general minimal residual method (GMRES)~\cite{saad1986gmres}. The LU decomposition in Step (1) and the backward and forward substitutions in Step (2c) are carried out using the sparse matrix direct solver PARDISO~\cite{alappat2020recursive}. The efficiency and the accuracy of this hybrid solver are demonstrated by the numerical experiments described in Section~\ref{sec:num_res}.


\subsection{Comments}\label{sec:comments}
Several comments about the formulation of the proposed HDG-BI solver detailed in Sections~\ref{sec:prob},~\ref{sec:disc},~\ref{sec:hdgbi},~\ref{sec:matrix}, and~\ref{sec:solver} are in order:
\begin{enumerate}
\item The proposed solver allows for the surface of the dielectric object (which is determined by $\varepsilon_{\mathrm{r}}$ and $\mu_{\mathrm{r}}$) to fully overlap with the computation domain boundary $\Gamma$. In such cases, the formulation detailed above stays the same without requiring any modifications. This type of flexibility is especially important for a concave scatterer since enforcing the BI equations on this type of scatterer's surface significantly reduces the size of the computation domain (compared to using ABCs or PML)~\cite{li2014hybrid, dong2022}.

\item The proposed solver can be easily modified to efficiently account for disconnected scatterers. In this case, the BI equations should be enforced separately on the surface of each scatterer. This approach eliminates the need to discretize the space around the scatterers resulting in a very efficient solver especially for scenarios where the scatterers are well separated (compared to using ABCs or PML which would require a computation domain that encloses all scatterers and call for its full discretization)~\cite{li2014hybrid, dong2022}.

\item Perfect electrically conducting (PEC) objects possibly present in the computation domain $\Omega$ can easily be accounted for with a small modification of the numerical flux described by~\eqref{eq:18a}. Assume that the triangular surface $S_l \in \partial \Omega_{i}$ has all its three corners on the PEC surface, then~\eqref{eq:18a} should be updated as~\cite{li2013hybridizable,li2014hybridizable}
\begin{equation}
\pi _{\partial {{\Omega}_{i}}}^{\times }\!\{ {{\mathbf{E}}_{\mathrm{h}}}\}+\pi _{\partial {{\Omega}_{i}}}^{\tau }\!\{{{\mathbf{H}}_{\mathrm{h}}}-{{\boldsymbol{\Lambda }}_{\mathrm{h}}}\}=0.\\
\label{eq:18c}
\end{equation}
\item The formulation described in this section assumes a conformal mesh, i.e., the triangular surfaces of any two neighboring tetrahedrons match (share the same three nodes) and similarly the triangular surfaces of the tetrahedrons match to those discretizing the computation domain boundary. The expansions of the local HDG unknowns $\mathbf{E}_{\mathrm{h}}$ and $\mathbf{H}_{\mathrm{h}}$, the global HDG unknown $\boldsymbol{\Lambda}_{\mathrm{h}}$, and the BI unknowns $\mathbf{J}_{\mathrm{h}}$ and $\mathbf{M}_{\mathrm{h}}$ are carried out independently and ``connected'' to each other using the numerical flux. Theoretically, this approach allows for non-conformal meshes to be used to discretize each of these sets of unknowns. Such an approach would require defining/computing the numerical flux on overlapping regions of non-matching triangular faces of different mesh sets. To the best of authors' knowledge, an HDG method, which can account for non-conformal meshes, has been developed for only 2D problems~\cite{solano2022,zhu2017}. The possibility of extending this method to 3D problems and to account for non-conformal surface meshes (for incorporation of BI) will be investigated in a future publication.

\item Electromagnetic scattering problems are often analyzed using integral equation solvers (for example see~\cite{peng2013discontinuous,sekulic2018versatile,lu2003fast}). Surface integral (SIE) solvers~\cite{peng2013discontinuous,sekulic2018versatile} when accelerated using MLFMA~\cite{rokhlin1990rapid,lu1994multilevel,song1997multilevel,sheng1998solution,fostier2008,ergul2008,michiels2015,yucel2018,abduljabbar2019} result in the most computationally efficient methods for scattering analysis. However, their applicability is limited to problems where the material properties are piecewise homogenous. In problems where the scatterer is inhomogeneous, one can switch to a volume integral equation (VIE) solver~\cite{lu2003fast}, but this type of solvers requires a volumetric discretization of the scatterer. The HDG-BI solver can set the computation domain boundary $\Gamma$ on the surface of the scatterer, ensuring that only the scatterer is discretized using a volumetric mesh. Under this condition, one can expect that the HDG-BI solver would be more efficient than the VIE solver. This is fundamentally because the volumetric discretization by the HDG-BI solver results in a sparse matrix while the volumetric discretization by the VIE solver results in a dense matrix. Indeed, the numerical results provided in Section~\ref{sec:layered} show that the proposed HDG-BI solver is faster that than a volume-surface integral equation (VSIE) solver in a problem where the scatterer is a PEC object embedded in a layered dielectric cube.

\end{enumerate}

\begin{figure}[t!]
\centering
\subfigure[]{\includegraphics[width=0.6\columnwidth,draft=false]{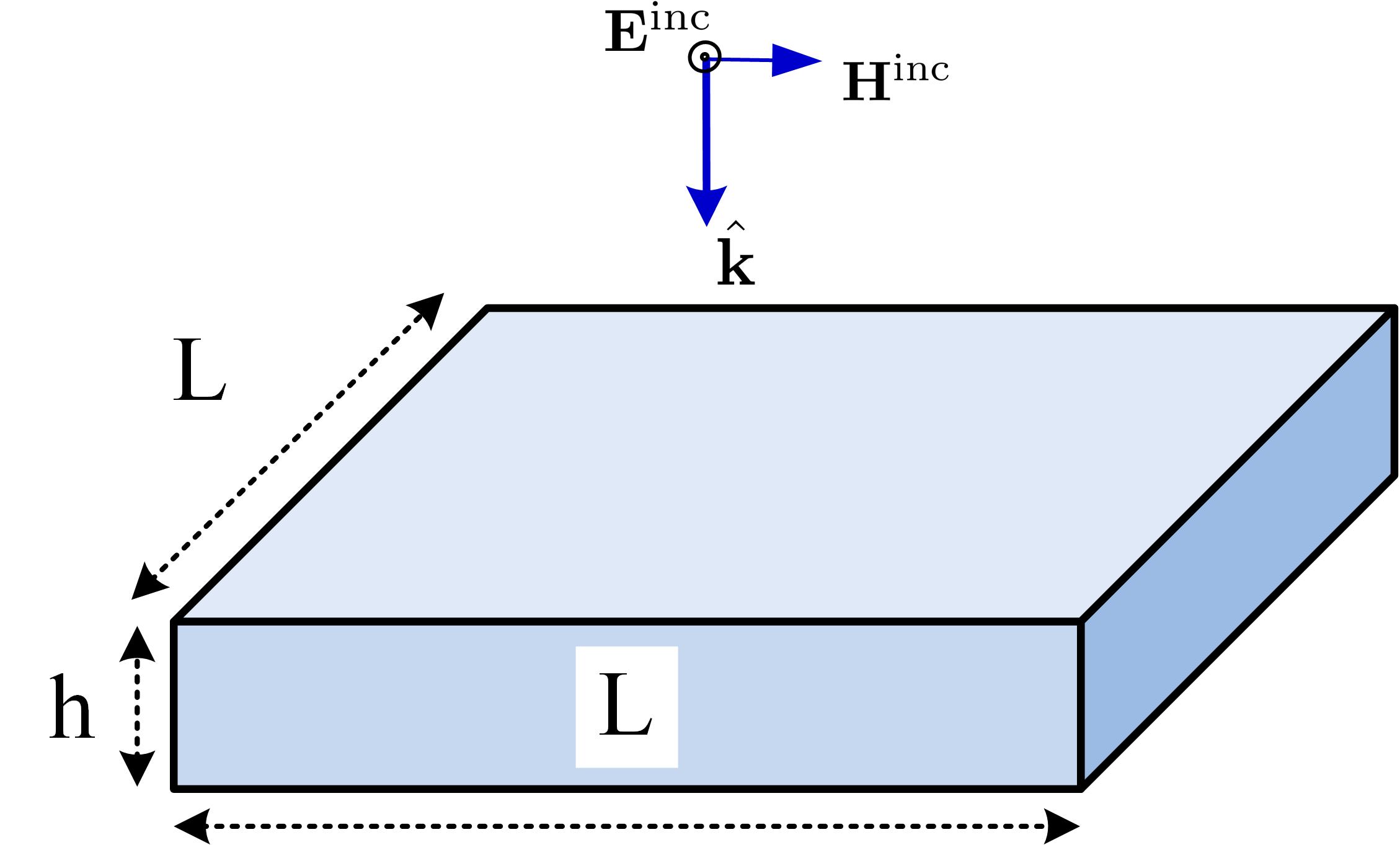}}
\subfigure[]{\includegraphics[width=0.6\columnwidth,draft=false]{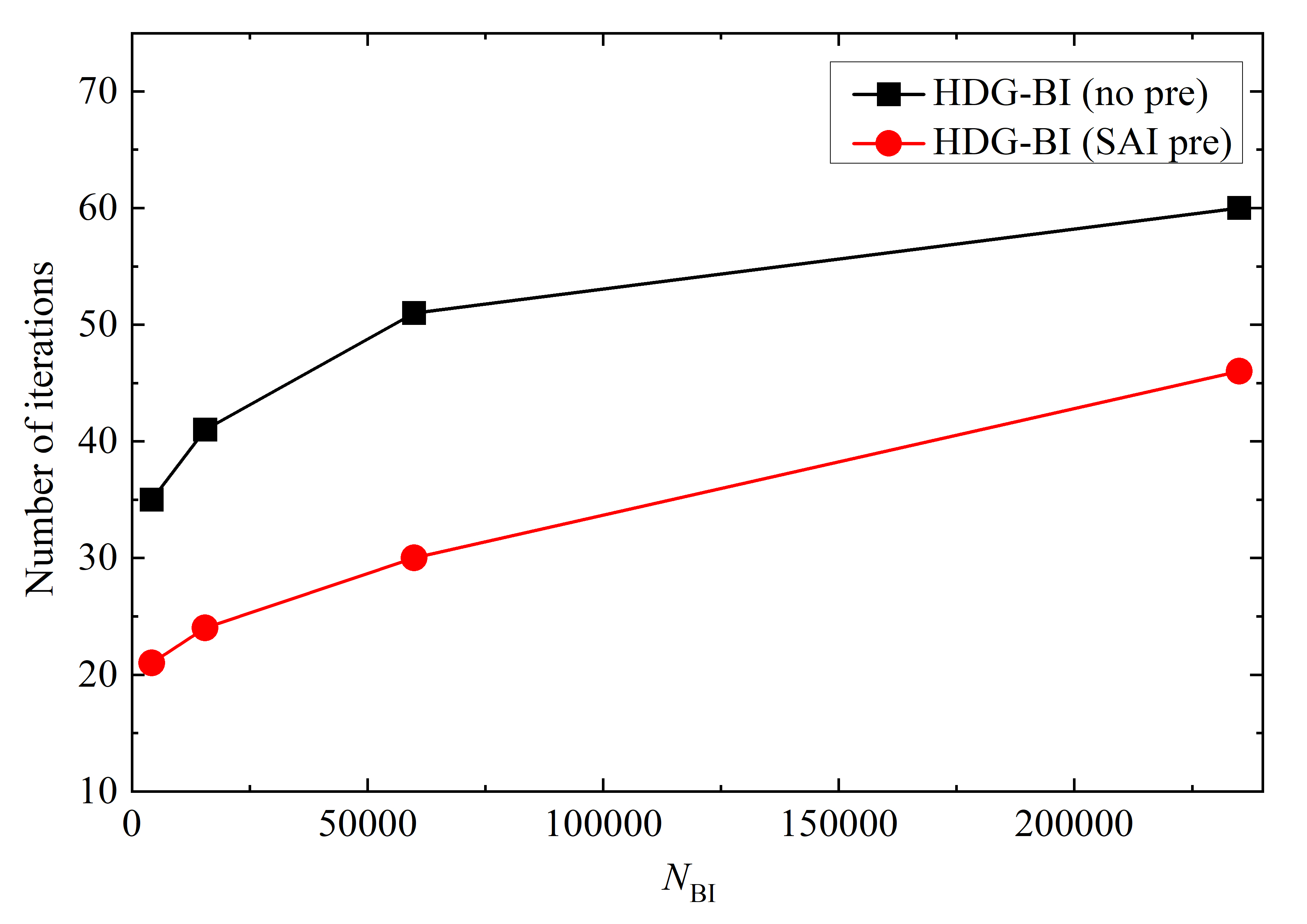}}
\caption{Electromagnetic scattering from a dielectric plate. (a) Description of the geometry and the excitation. (b) Number of iterations required by the GMRES method (without a preconditioner and with SAI preconditioner) versus $N_{\mathrm{BI}}$.}
\label{fig:plate}
\end{figure}

\section{Numerical Results}\label{sec:num_res}
In this section, several numerical examples are presented to demonstrate the accuracy, efficiency, and applicability of the proposed HDG-BI solver. In all examples, the scatterers are non-magnetic ($\mu_{\mathrm{r}}=1$ in the whole computation domain) and the background medium is the free space with permittivity $\varepsilon_0$ and permeability $\mu_0$. In all simulations, the excitation is a plane wave with electric and magnetic fields
\begin{equation}
    \label{eq:excitation}
    \begin{aligned}
    \mathbf{E}^{\mathrm{inc}}(\mathbf{r})&=E_0\mathbf{\hat p}e^{-jk_0\mathbf{\hat k}\cdot\mathbf{r}}\\
    \mathbf{H}^{\mathrm{inc}}(\mathbf{r})&=\frac{E_0}{\eta_0}\mathbf{\hat k}\times \mathbf{\hat p}e^{-jk_0\mathbf{\hat k}\cdot\mathbf{r}}
    \end{aligned}
\end{equation}
where $E_0=1\, \mathrm{V/m}$ is the electric field amplitude, $\mathbf{\hat p}$ is the unit vector along the direction of the electric field, $\mathbf{\hat k}$ is the unit vector along the direction of propagation, and $k_0=2\pi f/c_0$, $\eta_0=\sqrt{\mu_0/\epsilon_0}$, and $c_0$ are the wave number, impedance, and speed in the background medium, respectively. Here, $f$ is the frequency of excitation, and the wavelength in the background medium at this frequency is given by $\lambda_0=f/c_0$.

\subsection{ Dielectric Coated PEC Sphere}
In the first example, electromagnetic scattering from a dielectric coated PEC sphere is analyzed. The radius of the sphere is $0.3 \,\mathrm{m}$ and the thickness of the coating is $0.1\,\mathrm{m}$. The boundary of the computation domain (as denoted by $\Gamma$) is the outer surface of the coating. The excitation parameters are $f=0.3\,\mathrm{GHz}$, $\mathbf{\hat p}=\hat{\mathbf{x}}$, and $\mathbf{\hat k}=\hat{\mathbf{z}}$. A total of $12$ simulations are carried out using the HDG-BI solver for three different values of the coating's relative permittivity as $2.0$, $4.0$, and $8.0$ and four different discretizations of the computation domain with average edge lengths $0.1\lambda_0$, $0.075\lambda_0$, $0.05\lambda_0$, and $0.03\lambda_0$. Table~\ref{t:5-0} provides the values of $N_{\mathrm{HDG}}$ and $N_{\mathrm{BI}}$ (as used by the HDG-BI solver) and $N_{\mathrm{DG}}$ (as a reference) for these four different levels of discretization. In all simulations, the iterations of the GMRES method used in solving the matrix system~\eqref{eq:52} are terminated when the relative residual error reaches $0.001$. At the end of each simulation, the $L_2$-norm of the relative error in radar cross section (RCS) is computed using
\begin{equation}
error_{\sigma}=\sqrt{ \frac{\displaystyle \sum^N_{i=0} \left|\sigma(i\Delta\theta,\phi)-\sigma^\mathrm{ref}(i\Delta\theta,\phi)\right|^2}{\displaystyle \sum^N_{i=0} \left|\sigma^\mathrm{ref}(i\Delta\theta,\phi)\right|^2}}.
\label{eq:5-25}
\end{equation}
Here, $\sigma$ is the RCS computed using $\mathbf{J}$ and $\mathbf{M}$ obtained by the HDG-BI on $\Gamma$, $\sigma^\mathrm{ref}$ is the reference RCS computed using the Mie series solution, $N=180$, $\Delta \theta = 1.0 ^{\circ}$, and $\phi=0$.

Fig.~\ref{fig:rmseex1} plots $error_{\sigma}$ versus average element length for three different values of the coating's relative permittivity. As expected, the error decreases with the increasing mesh density regardless of the value of the coating's relative permittivity.
\begin{figure}
\centering
\subfigure[]{\includegraphics[width=0.6\columnwidth]{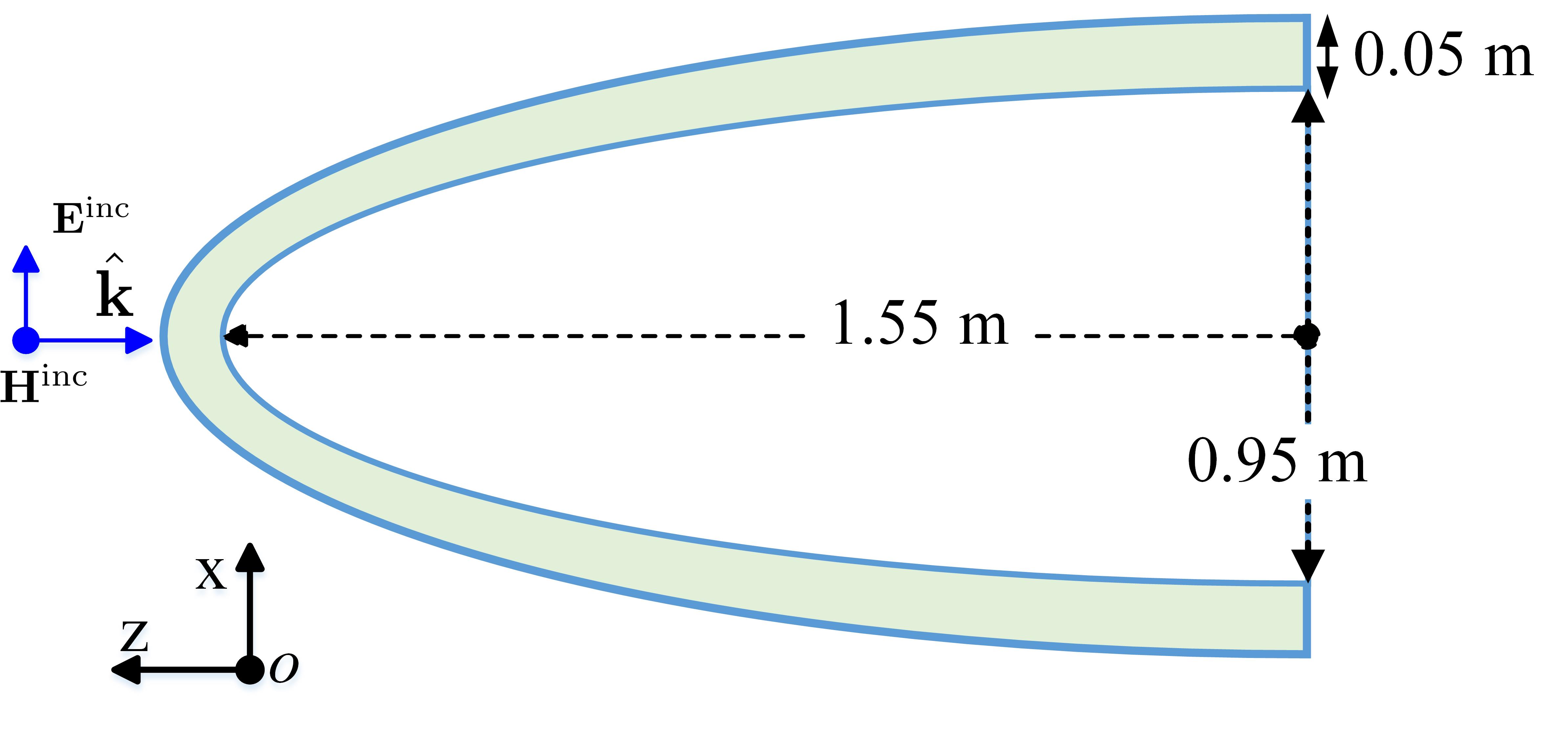}}\\
\subfigure[]{\includegraphics[width=0.46\columnwidth]{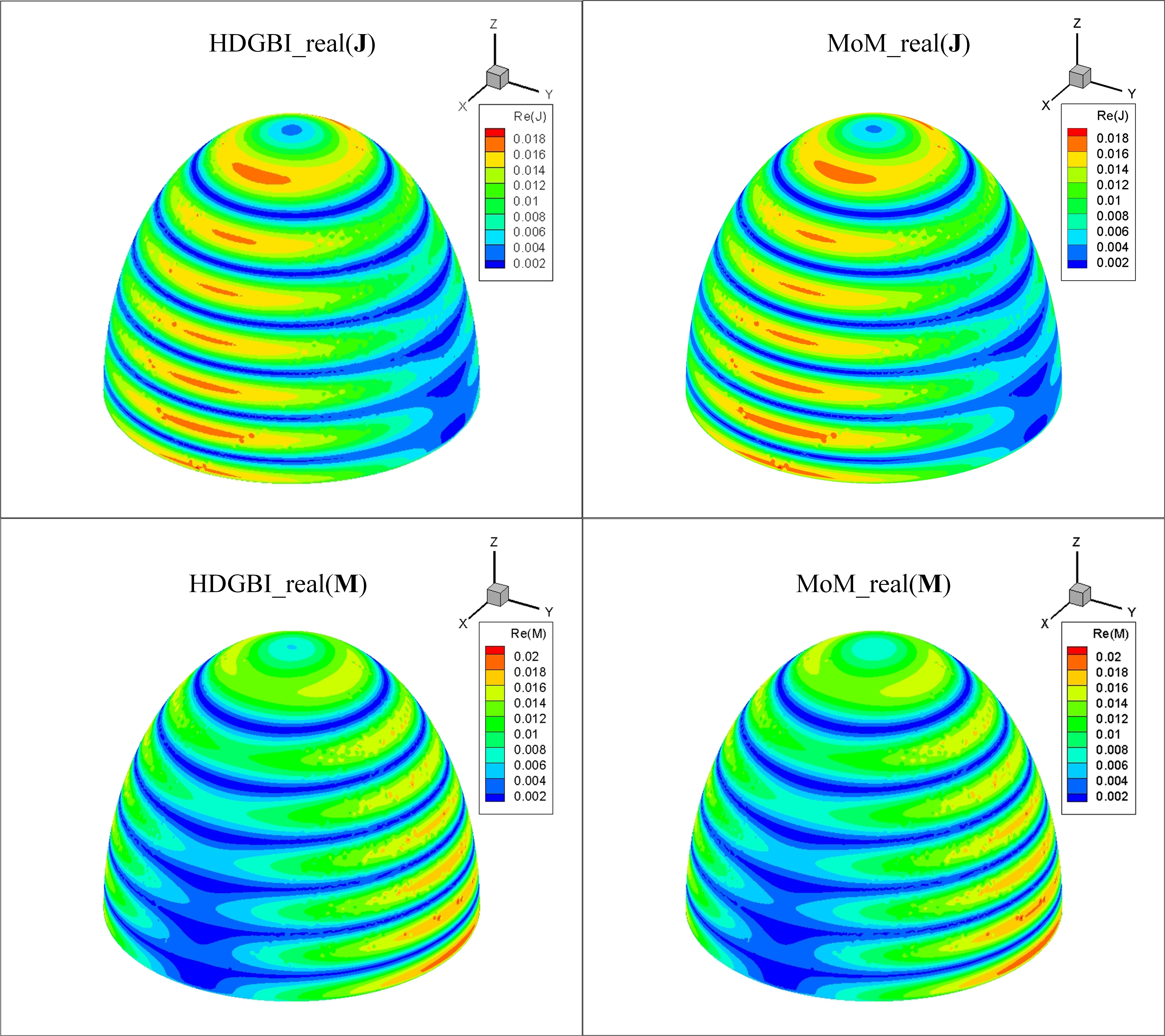}}
\subfigure[]{\includegraphics[width=0.46\columnwidth]{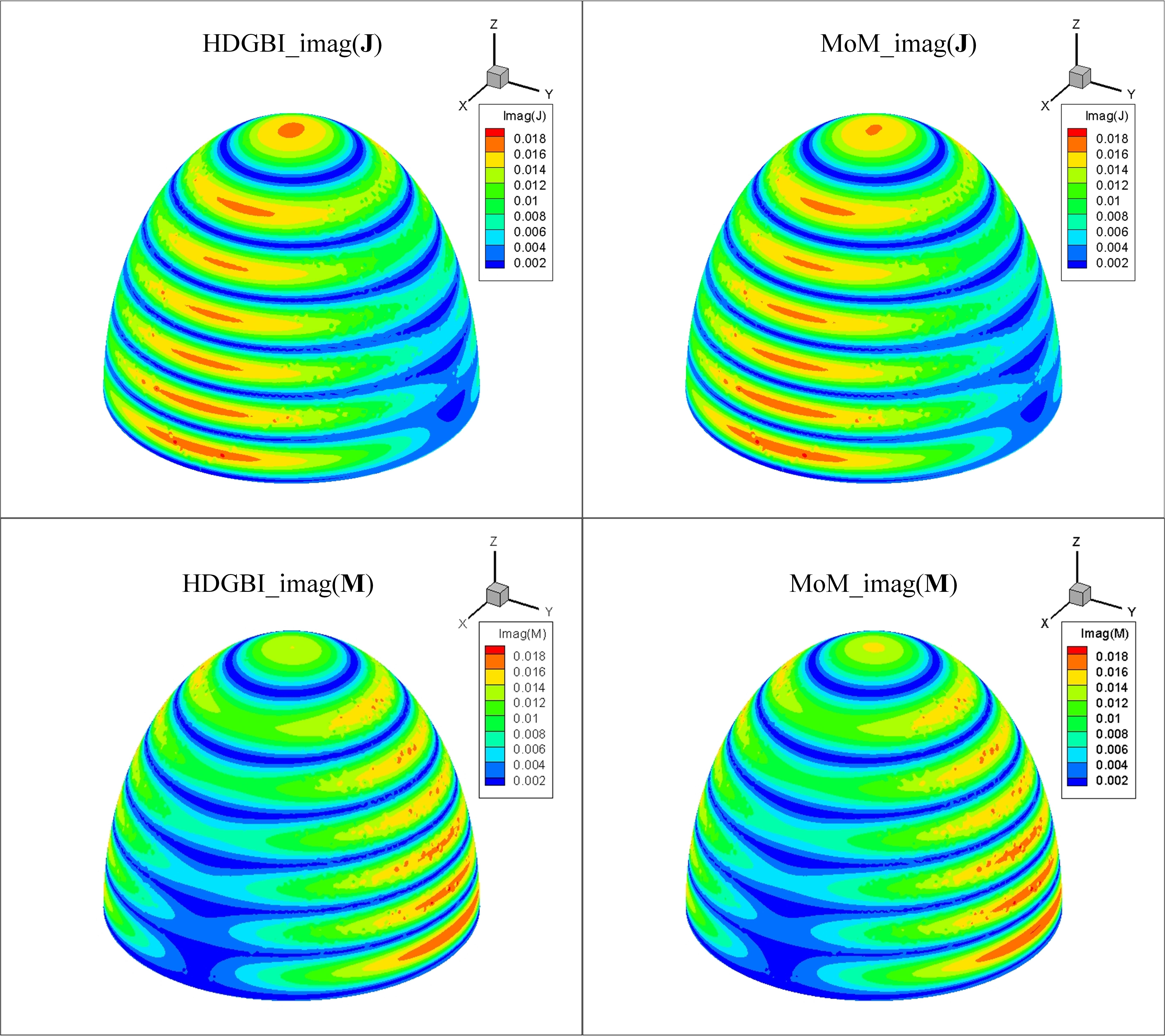}}\\\vspace{-0.25cm}
\subfigure[]{\includegraphics[width=0.6\columnwidth]{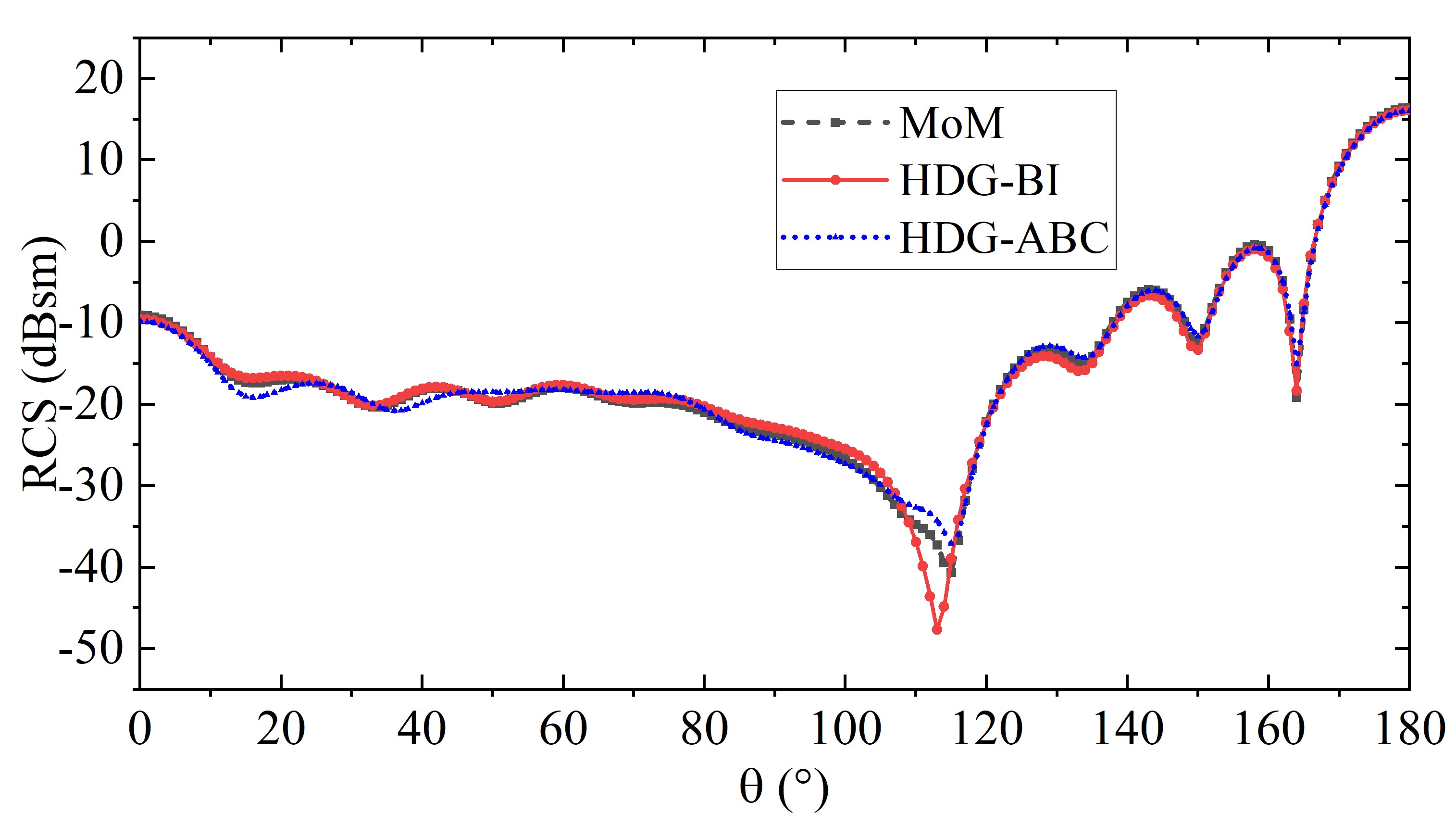}}\vspace{-0.25cm}
\caption{Electromagnetic scattering from a dielectric radome. (a) Description of the geometry (cross section) and the excitation. (b) Real and (c) imaginary part of $\mathbf{J}$ and $\mathbf{M}$ computed by HDG-BI and MoM on the surface of the radome. (d) RCS obtained using $\mathbf{J}$ and $\mathbf{M}$ that are computed by HDG-BI, HDG-ABC, and MoM.}
\label{fig:radome}
\end{figure}
\begin{figure*}
\centering
\subfigure[]{\includegraphics[width=0.6\columnwidth]{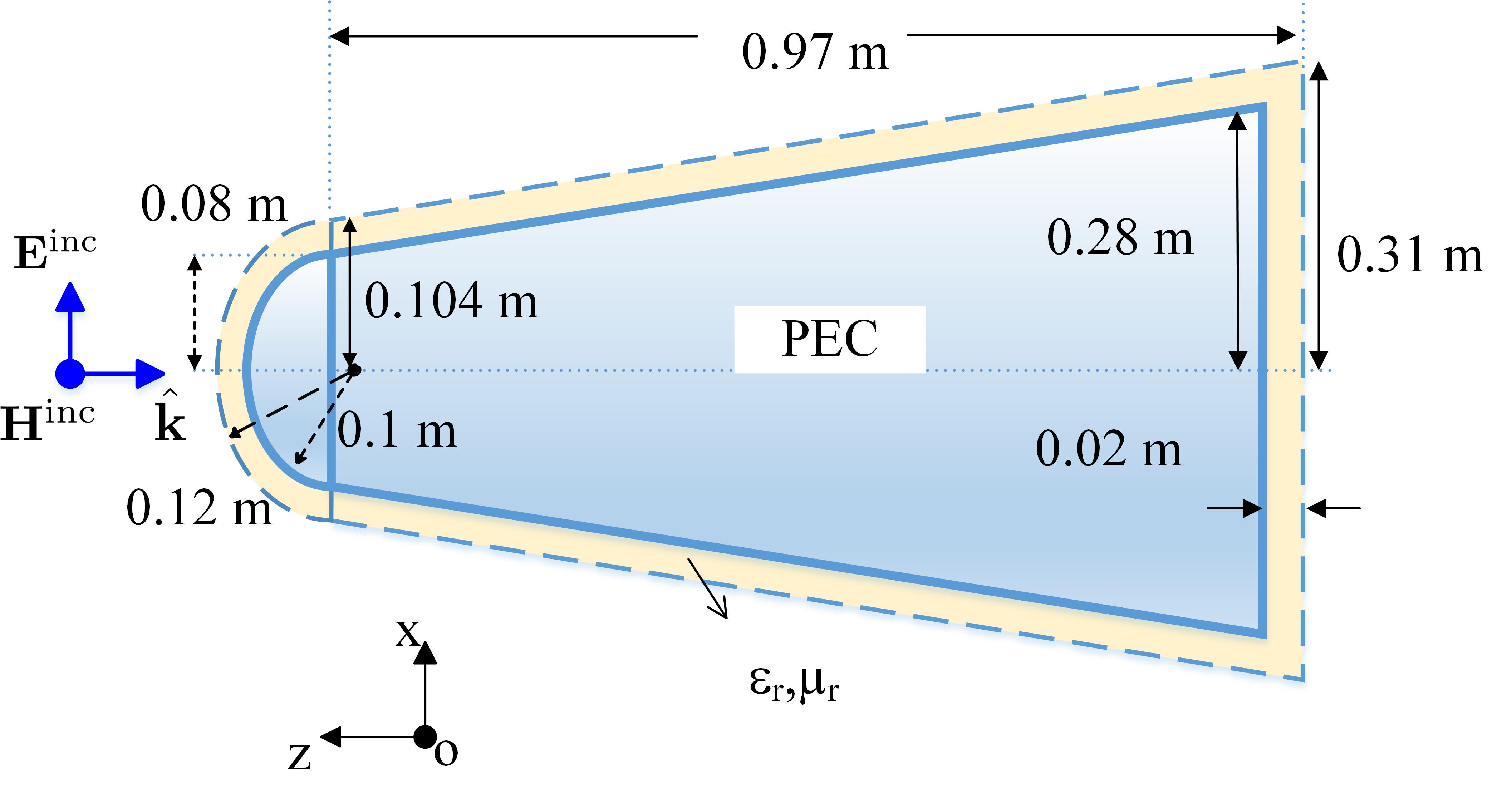}}\\
\subfigure[]{\includegraphics[width=0.46\columnwidth]{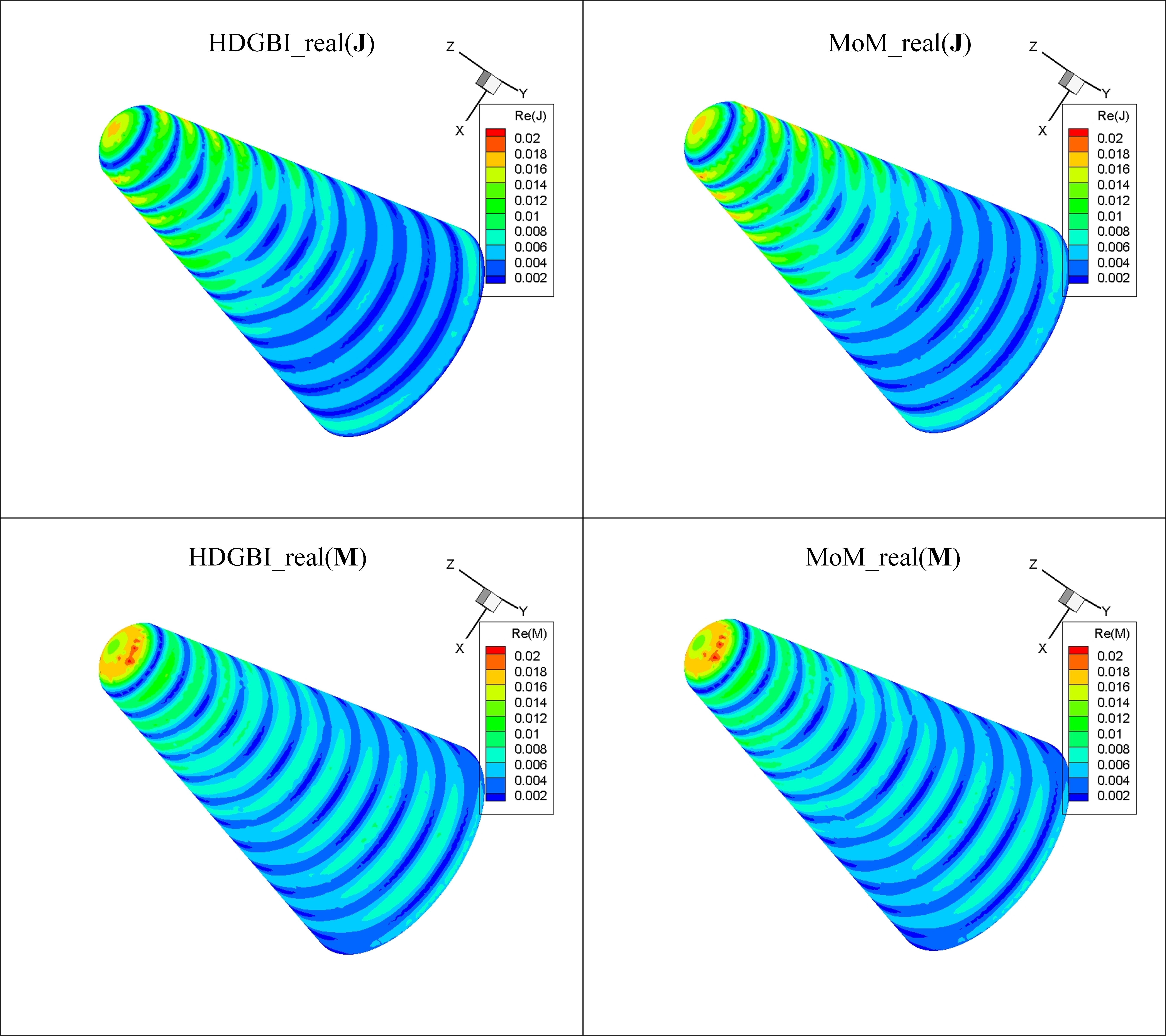}}
\subfigure[]{\includegraphics[width=0.46\columnwidth]{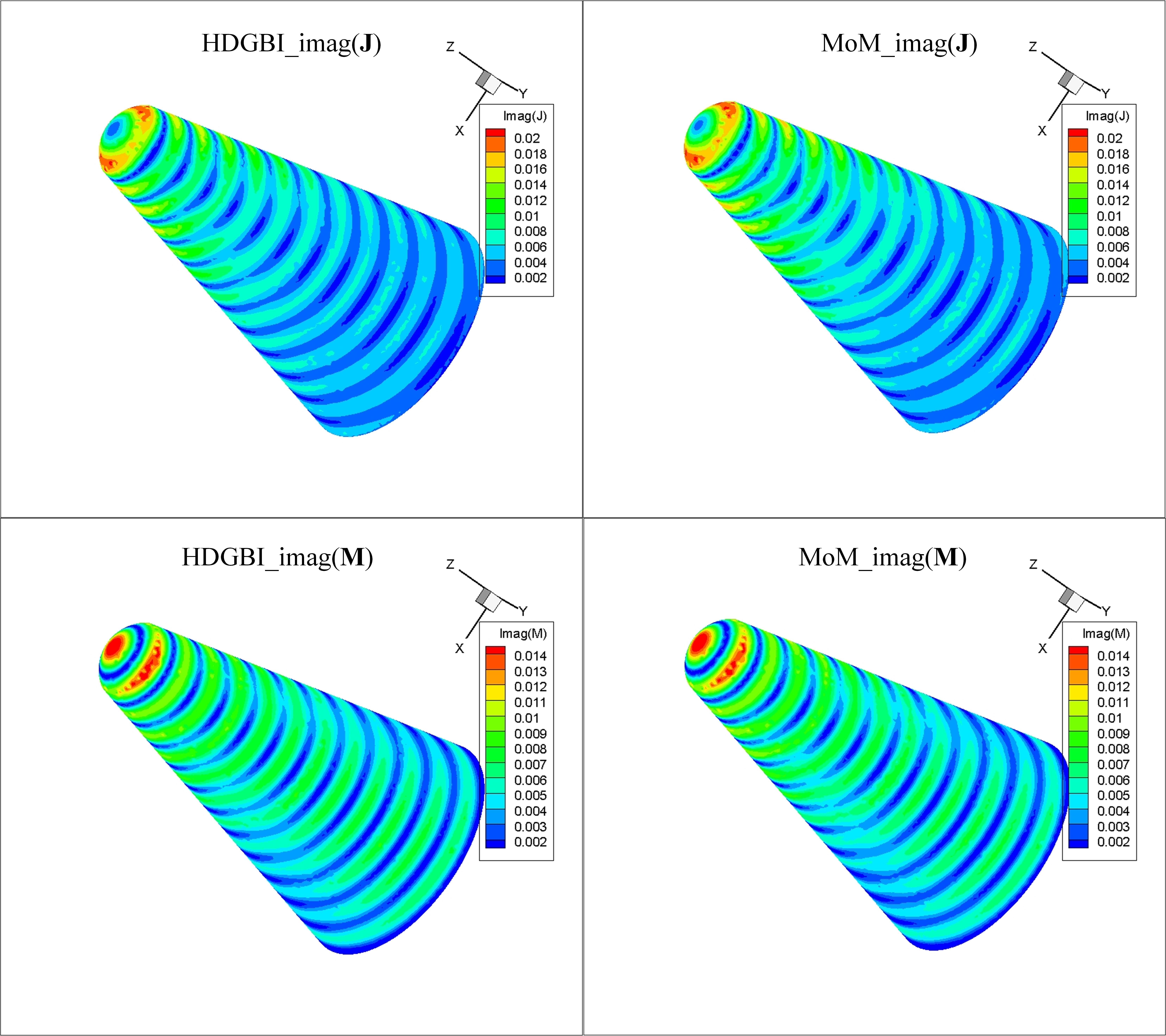}}\\\vspace{-0.25cm}
\subfigure[]{\includegraphics[width=0.46\columnwidth]{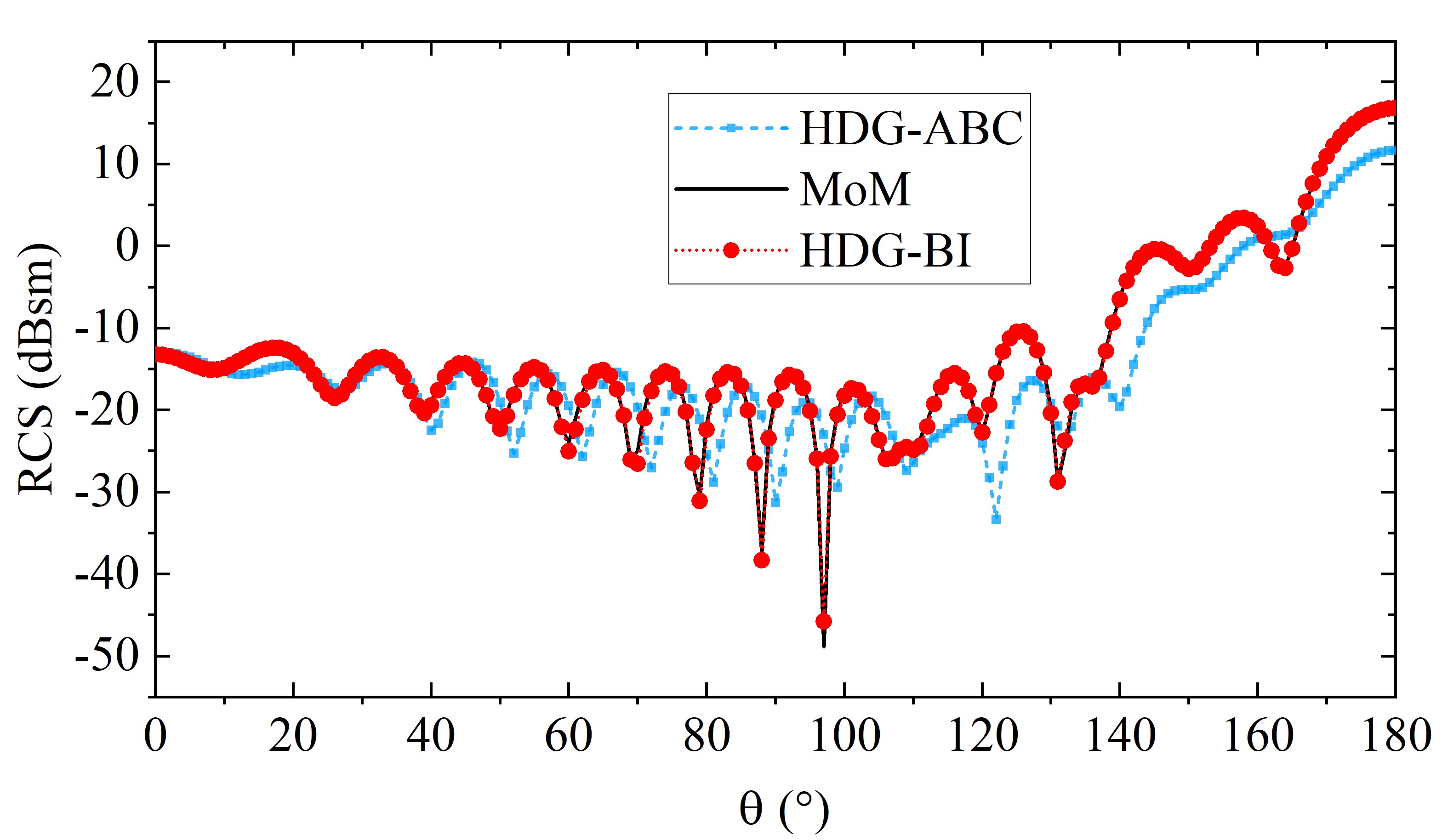}}
\subfigure[]{\includegraphics[width=0.46\columnwidth]{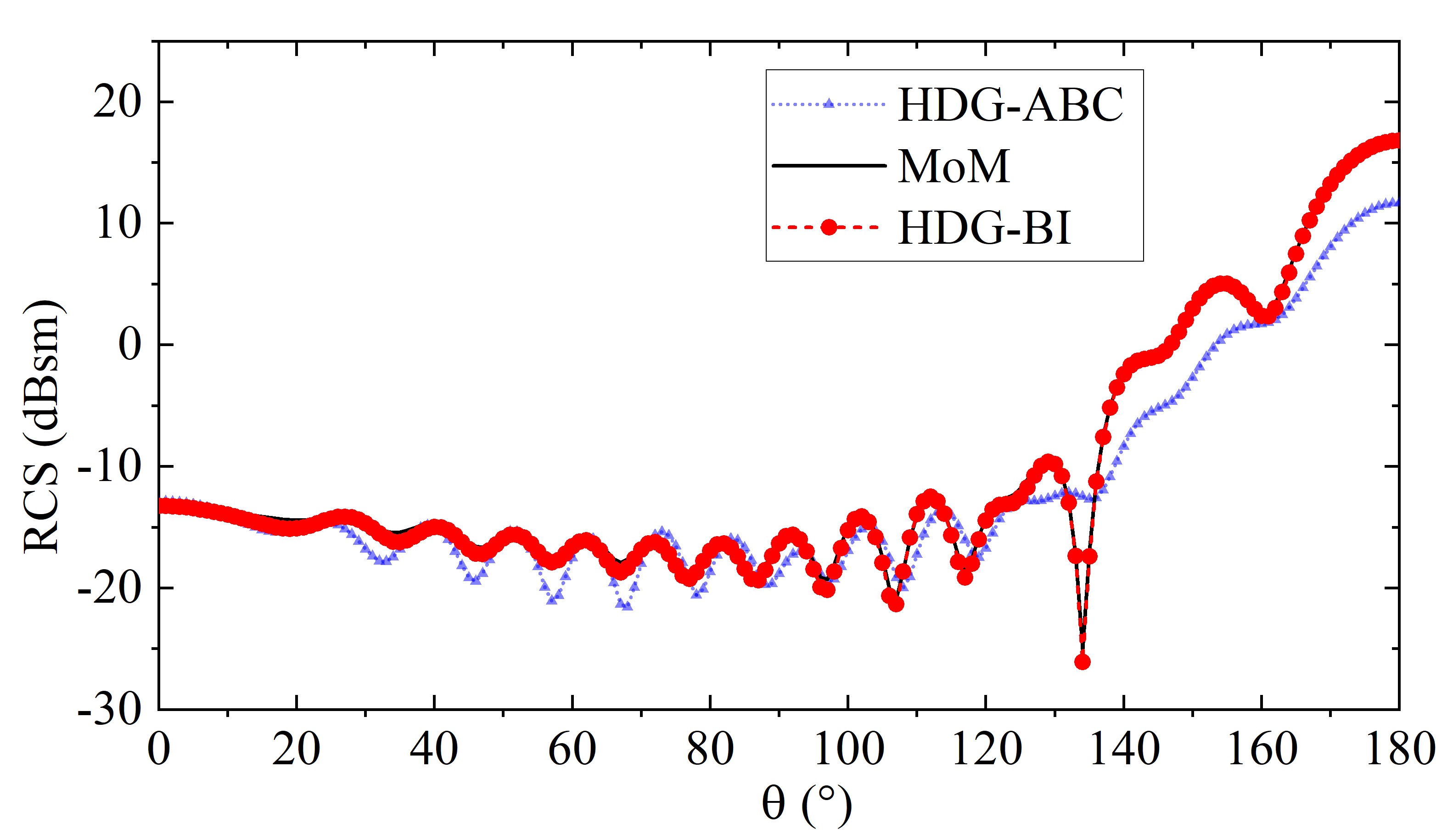}}\vspace{-0.25cm}
\caption{Electromagnetic scattering from a coated aircraft head. (a) Description of the geometry (cross section) and the excitation. (b) Real and (c) imaginary part of $\mathbf{J}$ and $\mathbf{M}$ computed by HDG-BI and MoM on the outer surface of the dielectric coating at for $\hat{\mathbf{p}}=\hat{\mathbf{x}}$ at $f=1.8\,\mathrm{GHz}$. RCS obtained using $\mathbf{J}$ and $\mathbf{M}$ that are computed by HDG-BI and MoM for (d) $\hat{\mathbf{p}}=\hat{\mathbf{x}}$ and (e) $\hat{\mathbf{p}}=\hat{\mathbf{y}}$ at  $f=1.8\,\mathrm{GHz}$.}
\label{fig:head}
\end{figure*}

\subsection{Dielectric Plate}
In the second example, electromagnetic scattering from a dielectric plate is analyzed. The dimensions of the plate are $L\times L \times h$ as shown in Fig.~\ref{fig:plate}(a), and its relative dielectric permittivity is $2.0$. The boundary of the computation domain (as denoted by $\Gamma$) is the surface of the plate. Four simulations are carried out for four different values of $L$, $L \in \{3.0, 6.0, 12.0, 24.0\}\, \mathrm{m}$. In all simulations, $h=0.1\,\mathrm{m}$ and the excitation parameters are  $f=0.3\,\mathrm{GHz}$, $\mathbf{\hat p}= \hat{\mathbf{x}}$, $\mathbf{\hat k}=-\hat{\mathbf{z}}$. The average edge lengths in the discretizations on the surface and in the volume of the plate are $0.1 \lambda_0$ and $0.075 \lambda_0$, respectively, resulting in $N_{\mathrm{HDG}} \in \{ 16\,468, 65\,168, 259\,965, 1\,037\,761\}$ and $N_{\mathrm{BI}} \in \{ 4\,174, 15\,550, 59\,870, 234\,990\}$ for four values of $L$. Two cases are considered for each simulation: (i) The matrix system~\eqref{eq:52} is solved using the GMRES method without using a preconditioner. (ii) The matrix system~\eqref{eq:52} is again solved using the GMRES method but this time a sparse approximate inverse (SAI) preconditioner~\cite{lee2004sparse,ibeid2018} is used. This preconditioner is constructed using only $\bar\bar{{C}}$. In both cases, the iterations of the GMRES method are terminated when the relative residual error reaches $0.001$.

Fig.~\ref{fig:plate}(b) plots the number of GMRES iterations versus $N_{\mathrm{BI}}$ for these two cases. For both cases, the slope of the iteration number curve flattens with increasing $N_{\mathrm{BI}}$, which means that the problem size does not have much effect on the efficacy of the iterative solver. Also, the figure shows that the SAI preconditioner can reduce the number of iterations. But it is worth mentioning here that for a more complicated scatterer (complex shape, inhomogeneous permittivity, etc.), this type of preconditioning may not be as effective~\cite{ibeid2018}.

\subsection{Dielectric Radome}
In this example, electromagnetic scattering from a dielectric radome is analyzed. Radome's cross section on the $xz$-plane is shown in Fig.~\ref{fig:radome}(a). The relative permittivity of the radome shell is $2.0$. The excitation parameters are $f=0.6\,\mathrm{GHz}$, $\mathbf{\hat p}= \hat{\mathbf{x}}$, and $\mathbf{\hat k}=-\hat{\mathbf{z}}$. Three simulations are carried out. (i) HDG-BI: The boundary of the computation domain (as denoted by $\Gamma$) is the surface the radome shell, i.e., HDG-BI discretizes only the volume of the shell and its surface. The average edge length in this discretization is $0.04 \lambda_0$ resulting in $N_{\mathrm{HDG}}=957\,468$ and $N_{\mathrm{BI}}=275\,604$. The iterations of the GMRES method used in solving the matrix system~\eqref{eq:52} are terminated when the relative residual error reaches $0.001$. (ii) HDG-ABC: The computation domain is a sphere of radius $1.5\,\mathrm{m}$. This sphere fully encloses the radome and the first-order ABC is enforced on its surface. The computation domain is discretized using elements with an average edge length of $0.04 \lambda_0$ resulting in $N_{\mathrm{HDG}}=4\,688\,763$. The HDG matrix system is solved using the sparse LU solver PARDISO~\cite{alappat2020recursive}. (iii) MoM: The multi-trace surface integral equation solver described in~\cite{zhao2020,zhao2021multitrace} is used.

Fig.~\ref{fig:radome}(b) and (c) compares the real and the imaginary parts of $\mathbf{J}$ and $\mathbf{M}$ obtained by HDG-BI and MoM solvers on the surface of the dielectric shell. The results agree well. RCS is computed for $\theta \in [0 \, 180^{\circ}]$ and $\phi=0$ using $\mathbf{J}$ and $\mathbf{M}$ obtained by the HDG-BI, HDG-ABC, and MoM solvers. Fig.~\ref{fig:radome}(d) plots RCS computed by these three solvers versus $\theta$ and shows that results obtained by the HDG-ABC and HDG-BI solvers agree well with those obtained by the MoM solver.

Even though both HDG-ABC and HDG-BI solvers produce accurate results for this problem, HDG-BI is significantly faster and has a much smaller memory imprint: The HDG-BI solver requires $9.16\,\mathrm{GB}$ of memory and completes the simulation in $1\,135\,\mathrm{s}$ while the HDG-ABC solver requires $118.4\,\mathrm{GB}$ of memory and completes the simulation in $3\,675\,\mathrm{s}$. Note that the number of GMRES iterations required by the HDG-BI solver is only $60$. This large difference in the computational requirements of these two solvers can be explained by the fact that the computation domain boundary where the ABC is enforced has to be located away from the randome surface to achieve the same accuracy level as the HDG-BI solver. This increases the computation domain size and, accordingly the computational requirements of the HDG-ABC solver.
\subsection{Aircraft Head}
\begin{figure}
\centering
\subfigure[]{\includegraphics[width=0.49\columnwidth]{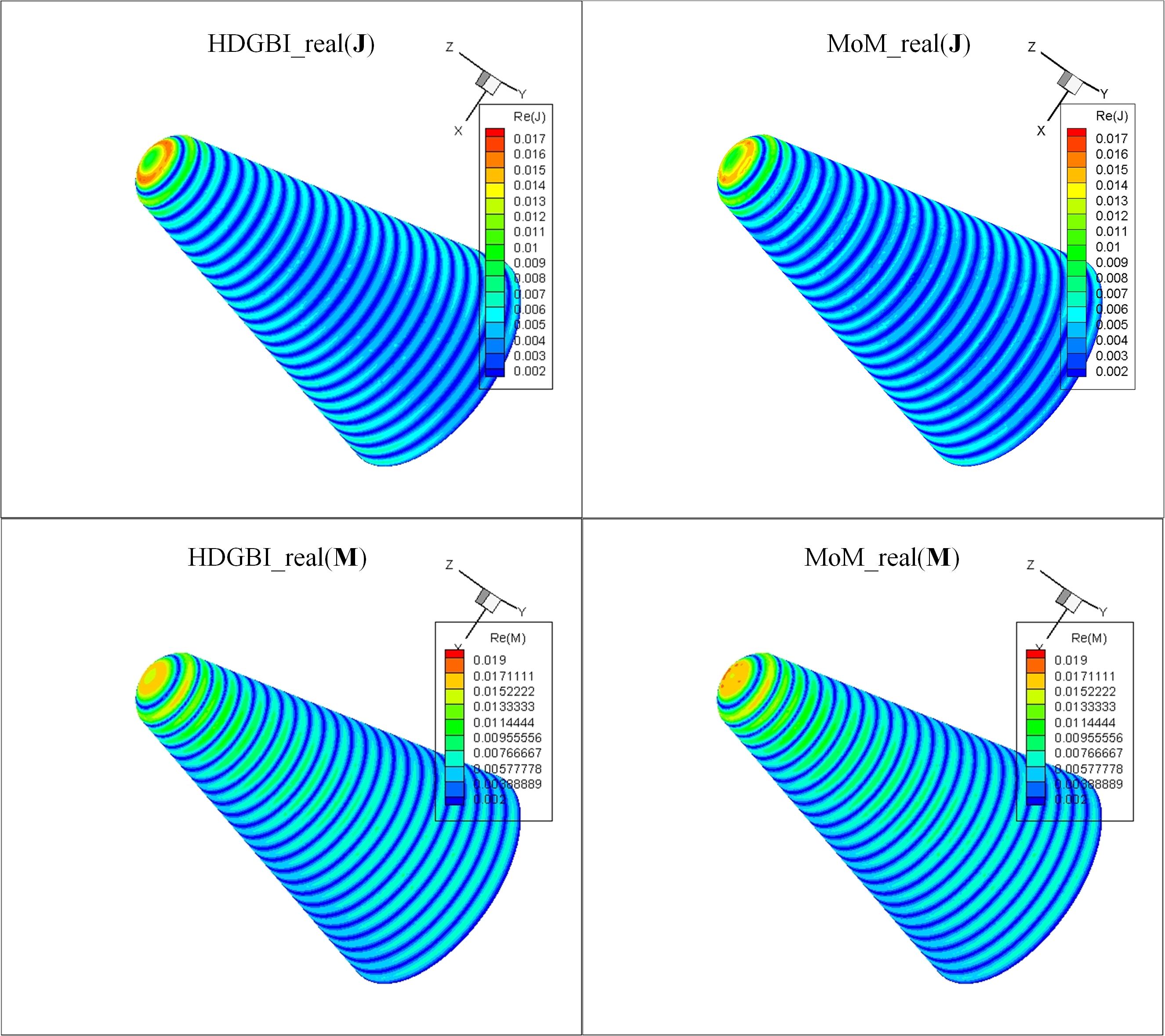}}
\subfigure[]{\includegraphics[width=0.49\columnwidth]{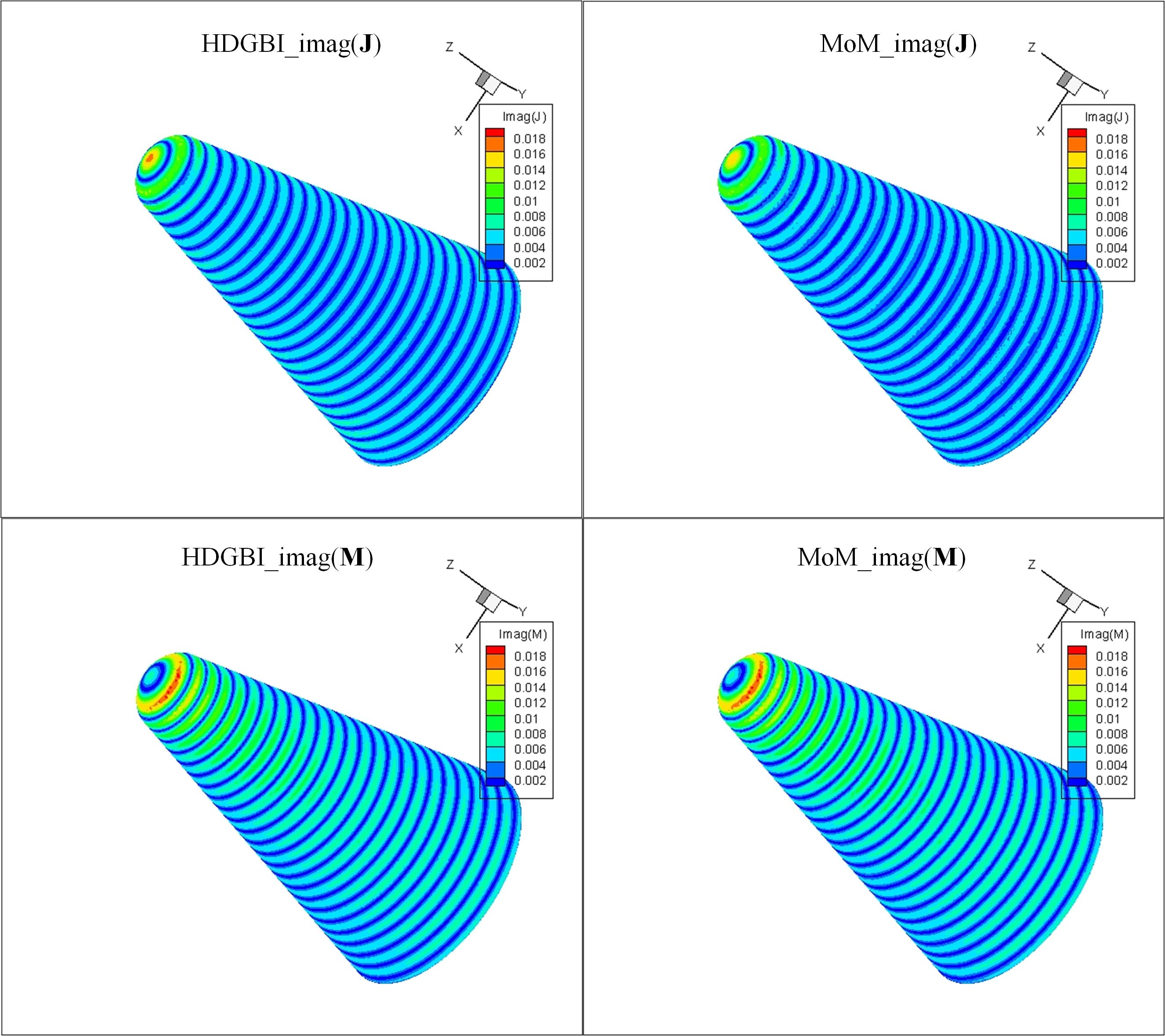}}
\caption{Electromagnetic scattering from an aircraft head. (a) Real and (b) imaginary part of $\mathbf{J}$ and $\mathbf{M}$ computed by HDG-BI and MoM on the outer surface of the dielectric coating for $\hat{\mathbf{p}}=\hat{\mathbf{x}}$ at $f=3.6\,\mathrm{GHz}$.}
\label{fig:head_36}
\end{figure}
In this example, electromagnetic scattering from a coated aircraft head model is analyzed. The aircraft head's cross section on the $xz$-plane is shown in Fig.~\ref{fig:head}. The relative permittivity of the coating is $2.0-0.5j$. The excitation parameters are $f \in \{1.8,3.6\}\,\mathrm{GHz}$, $\hat{\mathbf{k}}=-\hat{\mathbf{z}}$, and $\hat{\mathbf{p}} \in \{\mathbf{x},\mathbf{y}\}$. Three sets of simulations are carried out. (i) HDG-BI: The boundary of the computation domain (as denoted by $\Gamma$) is the outer surface of the coating, i.e., HGD-BI discretizes only the volume of the coating and its surface. Two levels of discretization with average edge lengths $0.07\lambda_0$ (resulting in $N_{\mathrm{HDG}}=813\,357$ and $N_{\mathrm{BI}}=81\,192$) and $0.075\lambda_0$ (resulting in $N_{\mathrm{HDG}}=8\,696\,010$ and $N_{\mathrm{BI}}=316\,086$) are used for the simulations with $f=1.8\,\mathrm{GHz}$ and $f=3.6\,\mathrm{GHz}$, respectively. The iterations of the GMRES method used in solving the matrix system~\eqref{eq:52} are terminated when the relative residual error reaches $0.001$. (ii) HDG-ABC: The computation domain is a sphere of radius $0.8\,\mathrm{m}$. This sphere fully encloses the coated aircraft head and the first-order ABC is enforced on its surface. The computation domain is discretized using elements with an average edge length of $0.07 \lambda_0$ resulting in $N_{\mathrm{HDG}}=8\,505\,207$ for the simulation with $f=1.8\,\mathrm{GHz}$. The HDG matrix system is solved using the sparse LU solver PARDISO~\cite{alappat2020recursive}. Note that for the simulation with $f=3.6\,\mathrm{GHz}$, the HDG-ABC solver is not used because of its prohibitive computational requirements. (iii) MoM: The multi-trace surface integral equation solver described in~\cite{zhao2020,zhao2021multitrace} is used.

Fig.~\ref{fig:head} (b) and (c) compare the real and the imaginary of parts of $\mathbf{J}$ and $\mathbf{M}$ obtained by HDG-BI and MoM solvers on the outer surface of the dielectric coating for the simulation with $f=1.8{\mathrm{GHz}}$ and $\hat{\mathbf{p}}=\hat{\mathbf{x}}$. The results agree well. RCS is computed for $\theta \in [0 \, 180^{\circ}]$ and $\phi=0$ using $\mathbf{J}$ and $\mathbf{M}$ obtained by the HDG-BI, HDG-ABC, and MoM solvers in two simulations with $\hat{\mathbf{p}}=\hat{\mathbf{x}}$ and $\hat{\mathbf{p}}=\hat{\mathbf{y}}$. Fig.~\ref{fig:head} (d) and (e) plots RCS computed by these three solvers versus $\theta$ in the simulations with $\hat{\mathbf{p}}=\hat{\mathbf{x}}$ and $\hat{\mathbf{p}}=\hat{\mathbf{y}}$, respectively. The figure shows that the results obtained by the HDG-BI solver agree very well with those obtained by the MoM solver while the results obtained by the HDG-ABC solver do not. The inaccuracy of the HDG-ABC solver can be explained by the fact that the first-order ABC is used to truncate the computation domain. As expected, the HDG-BI solver does not suffer from this bottleneck.

Furthermore, the computational requirements of the HDG-BI solver are significantly lower than those of the HDG-ABC solver: The HDG-BI solver requires $6.7\,\mathrm{GB}$ of memory and completes the simulation in $165\,\mathrm{s}$ while the HDB-ABC solver requires $210.1\,\mathrm{GB}$ of memory and completes the simulation in $3.58\,\mathrm{h}$. Note that the number of GMRES iterations required by the HDG-BI solver is only $45$. The large difference in the computational requirements of these two solvers can be explained by the fact that the computation domain of the HDG-ABC and the degrees of freedom required its discretization (as represented by $N_{\mathrm{HDG}}$) are significantly larger than those of the HDG-BI solver.

Finally, Fig.~\ref{fig:head_36} (a) and (b) compare the real and the imaginary of parts of $\mathbf{J}$ and $\mathbf{M}$ obtained by HDG-BI and MoM solvers on the outer surface of the dielectric coating for the simulation with $f=3.6\,{\mathrm{GHz}}$ and $\hat{\mathbf{p}}=\hat{\mathbf{x}}$. The results agree well. The HDG-BI solver completes this simulation in $1\,993\,\mathrm{s}$ and requires $99.5\,\mathrm{GB}$ of memory. The number of GMRES iterations is only $35$.

\begin{figure}[t!]
\centering
\subfigure[]{\includegraphics[width=0.6\columnwidth]{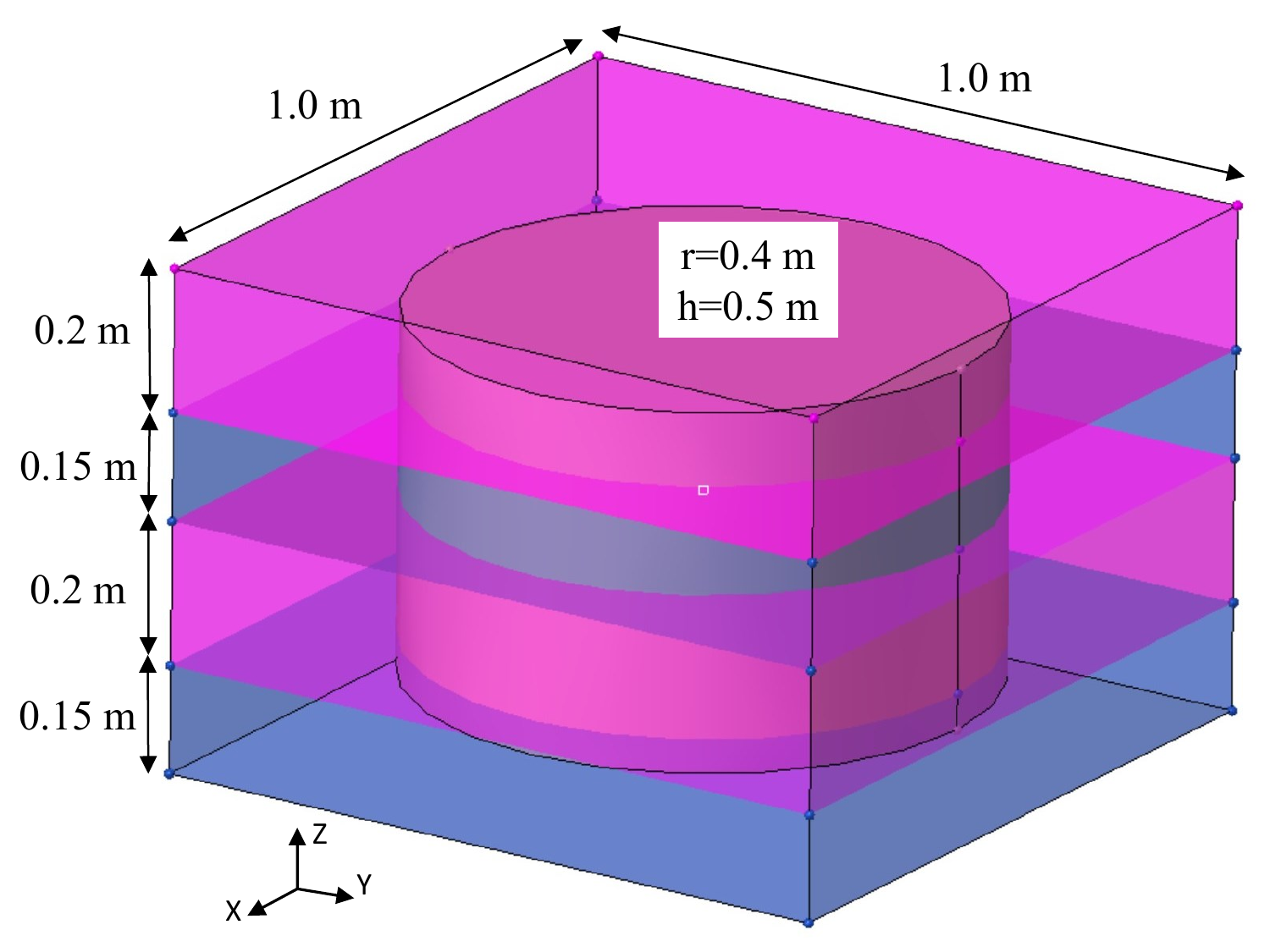}}
\subfigure[]{\includegraphics[width=0.6\columnwidth]{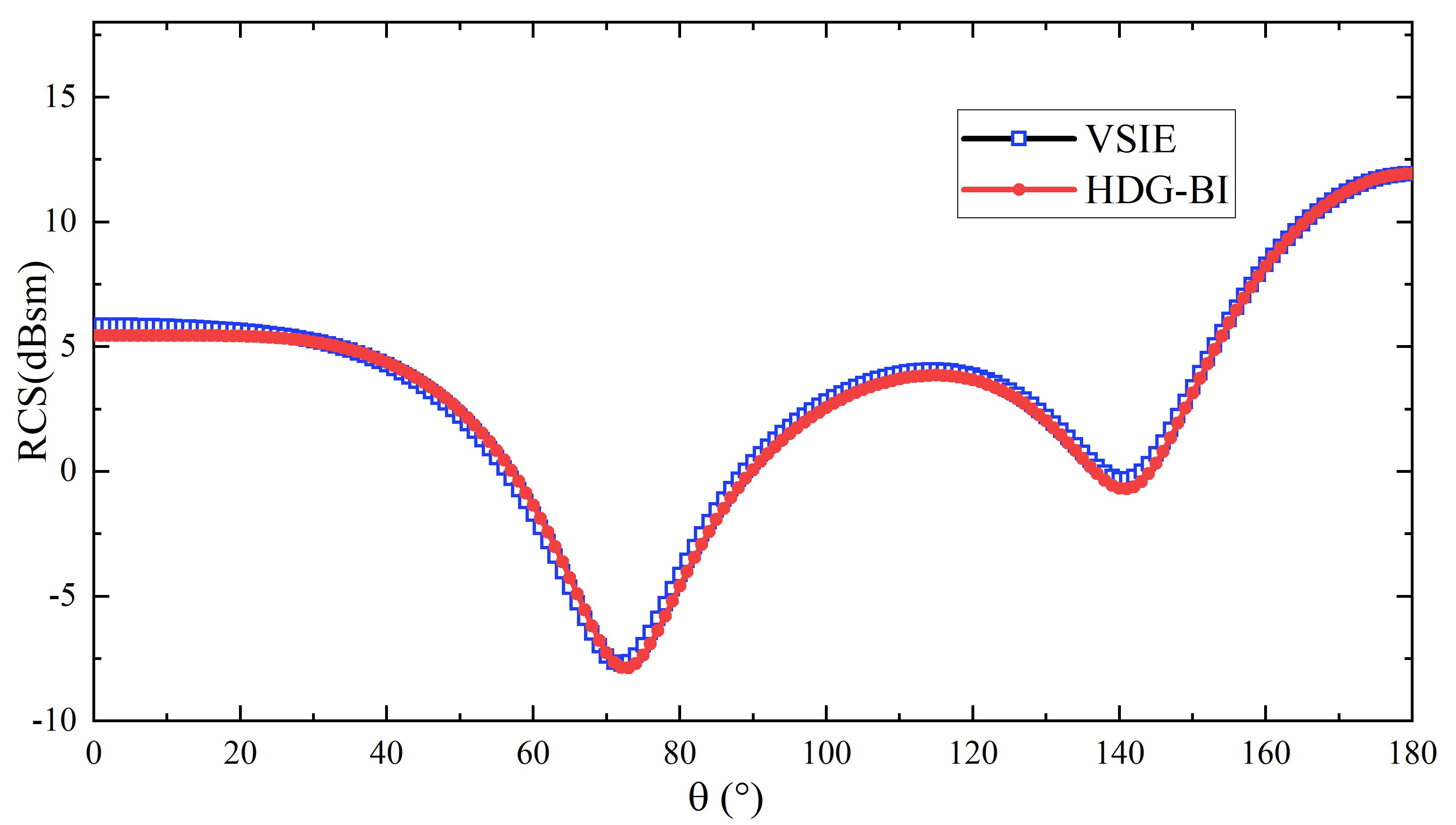}}
\caption{Electromagnetic scattering from a PEC cylinder embedded in a layered dielectric cube. (a) Description of the geometry (b) RCS computed using the HDG-BI and the VSIE solvers.}
\label{fig:ingeo}
\end{figure}

\subsection{PEC Cylinder Embedded in a Layered Dielectric Cube}\label{sec:layered}

In the last example, electromagnetic scattering from a PEC cylinder embedded in a layered dielectric cube is analyzed. The geometry of the scatterer is shown in Fig.~\ref{fig:ingeo} (a). The relative permittivities of the four layers (ordered from top to bottom) are $3.0$, $2.0$, $3.0$, and $2.0$. The excitation parameters are $f=0.3\,\mathrm{GHz}$, $\mathbf{\hat p}= \hat{\mathbf{x}}$, and $\mathbf{\hat k}=-\hat{\mathbf{z}}$. Two simulations are carried out: (i) HDG-BI: The boundary of the computation domain (as denoted by $\Gamma$) is the surface of the cylinder and the outer surface of the cube. The computation domain is discretized using elements with an average edge length of $0.05 \lambda_0$ resulting $N_{\mathrm{HDG}}=201\,294$ and $N_{\mathrm{BI}}=11\,856$. The iterations of the GMRES method used in solving the matrix system~\eqref{eq:52} are terminated when the relative residual error reaches $0.001$. (ii) VSIE: The commercially available software package FEKO~\cite{feko} that solves a coupled system of VIE (enforced inside the cube) and SIE (enforced on the surface of the cylinder) is used. The average edge length in the software is set to $0.05 \lambda_0$ which results in $74\,162$ degrees of freedom for the VSIE solver. Note that neither the HDG-BI solver nor the VSIE solver is accelerated using MLFMA.

Fig.~\ref{fig:ingeo} (b) plots RCS computed for $\theta \in [0 \, 180^{\circ}]$ and $\phi=0$ in these two simulations. Results agree well demonstrating the accuracy of the proposed HDG-BI solver. For this problem, the computational requirements of the HDG-BI solver are significantly lower than those of the VSIE solver: The HDG-BI solver requires $2.48\,\mathrm{GB}$ of memory and completes the simulation in $13.35\,\mathrm{m}$ ($8.03\,\mathrm{m}$ to compute the matrix and $5.32\,\mathrm{m}$ to solve the matrix system) while the VSIE solver requires $41.55\,\mathrm{GB}$ of memory and completes the simulation in $77.72\,\mathrm{m}$ ($18.12\,\mathrm{m}$ to compute the matrix and $59.6\,\mathrm{m}$ to solve the matrix system). This comparison shows the benefits of the proposed HDG-BI solver over the VSIE solver, which mainly stems from the fact that the volumetric discretization by HDG results in a sparse matrix while the volumetric discretization by VSIE results in a dense matrix. 

\section{Conclusions}

A method, which couples the HDG and the BI equations is developed to efficiently analyze electromagnetic scattering from inhomogeneous/composite objects. The coupling between these two sets of equations is realized using the numerical flux operating on the equivalent current and the global unknown of the HDG. This approach yields sparse coupling matrices upon discretization. Inclusion of the BI equation ensures that the only error in enforcing the radiation conditions is the discretization. Furthermore, the computation domain boundary, where the BI equation is enforced, can be located very close, even conformal, to the surface of the scatterer without any loss of accuracy. This significantly reduces the number of unknowns to be solved for compared to the traditional HDG schemes that make use of ABCs or PML to truncate the computation domain.

However, the discretization of the BI equation yields a dense matrix, which prohibits the use of a direct matrix solver on the overall coupled system as often done with traditional HDG schemes. To overcome this bottleneck, a ``hybrid'' method is developed. This method uses an iterative scheme to solve the overall coupled system but within the matrix-vector multiplication subroutine of the iterations, the inverse of the HDG matrix is efficiently accounted for using a sparse direct matrix solver. The same subroutine also uses the multilevel fast multipole algorithm to accelerate the multiplication of the guess vector with the dense BI matrix.  Numerical examples show that the proposed HDG-BI solver has clear advantages over the traditional HDG schemes with ABCs and a VSIE solver.

As future work, a domain decomposition method and high-order vector basis functions will be incorporated into the HDG-BI solver to further improve its efficiency and accuracy and its applicability to large-scale problems. Additionally, a discretization scheme that can account for non-conformal meshes will be formulated and implemented within the framework of the HDG-BI solver.



\end{document}